\documentclass[10pt,conference,dvipsnames]{IEEEtran}
\IEEEoverridecommandlockouts
\usepackage{cite}
\usepackage{textcomp}
\usepackage{xcolor}
\def\BibTeX{{\rm B\kern-.05em{\sc i\kern-.025em b}\kern-.08em
    T\kern-.1667em\lower.7ex\hbox{E}\kern-.125emX}}

\usepackage[utf8]{inputenc}
\usepackage[T1]{fontenc}
\usepackage{amsmath,amssymb,amsbsy}
\usepackage{mathtools}
\usepackage{xparse}
\usepackage{todonotes}
\presetkeys{todonotes}{inline}{}
\usepackage{comment}

\usepackage{graphicx}
\usepackage{dcolumn}
\usepackage{bm}
\usepackage{lipsum}
\usepackage{amsmath}
\usepackage{amssymb}
\usepackage{float}
\usepackage{setspace}

\usepackage[normalem]{ulem}
\usepackage{xcolor}
\usepackage{caption,subcaption}
\usepackage{comment}
\usepackage{array,multirow,booktabs}
\usepackage[makeroom]{cancel}
\usepackage{amsthm}
\usepackage{mathtools}
\usepackage{comment}
\usepackage{standalone}
\usepackage{multicol}

\usepackage{adjustbox}
\usepackage{tikz}
\usetikzlibrary{quantikz}
\usetikzlibrary{external}

\usepackage{cprotect}

\usepackage[colorlinks=true]{hyperref}
\usepackage{doi}

\usepackage{enumitem}

\theoremstyle{definition}

\newcommand{\Id}{\mathit{Id}}
\newcommand{\CNOT}{\mathit{CNOT}}
\newcommand{\SWAP}{\mathit{SWAP}}

\newcommand{\pqbit}{\mathit{pq}}
\newcommand{\lqbit}{\mathit{lq}}
\newcommand{\anc}{\mathit{an}}

\NewDocumentCommand{\bbeta}{o}{
	\pmb{\beta}\IfValueT{#1}{_{[#1]}}
}
\NewDocumentCommand{\bgamma}{o}{
	\pmb{\gamma}\IfValueT{#1}{_{[#1]}}
}
\NewDocumentCommand{\bgstate}{o}{
	\ket{\IfValueTF{#1}{\bbeta[#1]}{\bbeta},
		\IfValueTF{#1}{\bgamma[#1]}{\bgamma}
	}
}
\NewDocumentCommand{\bgstateT}{o}{
	\bra{	\IfValueTF{#1}{\bbeta[#1]}{\bbeta},
		\IfValueTF{#1}{\bgamma[#1]}{\bgamma}
	}
}

\title{Sampling on NISQ Devices: \\ ``Who's the Fairest One of All?'' %

\thanks{Research presented in this article was supported by the Laboratory Directed Research and Development program of Los Alamos National Laboratory under project number 20200671DI.
\hfill LA-UR-21-25803
}}

\author{
    \IEEEauthorblockN{Elijah Pelofske\IEEEauthorrefmark{1}, 
    John Golden\IEEEauthorrefmark{1}, 
    Andreas Bärtschi\IEEEauthorrefmark{1}, 
    Daniel O'Malley\IEEEauthorrefmark{2} and 
    Stephan Eidenbenz\IEEEauthorrefmark{1}}
    \IEEEauthorblockA{\IEEEauthorrefmark{1}
    \textit{CCS-3 Information Sciences, Los Alamos National Laboratory},
	Los Alamos, NM 87544, USA \\
	Email: epelofske@lanl.gov, golden@lanl.gov, baertschi@lanl.gov, eidenben@lanl.gov
    }
    \IEEEauthorblockA{\IEEEauthorrefmark{2}
    \textit{EES-16 Earth Sciences, Los Alamos National Laboratory},
	Los Alamos, NM 87544, USA \\
	Email: omalled@lanl.gov
    }
}

\begin{document}

\maketitle

\begin{abstract}
Modern NISQ devices are subject to a variety of biases and sources of noise that degrade the solution quality of computations carried out on these devices. A natural question that arises in the NISQ era, is how fairly do these devices sample ground state solutions. 
To this end, we run five fair sampling problems (each with at least three ground state solutions) that are based both on quantum annealing and on the Grover Mixer-QAOA algorithm for gate-based NISQ hardware. In particular, we use seven IBM~Q devices, the Aspen-9 Rigetti device, the IonQ device, and three D-Wave quantum annealers.

For each of the fair sampling problems, we measure the ground state probability, the relative fairness of the frequency of each ground state solution with respect to the other ground state solutions, and the aggregate error as given by each hardware provider. Overall, our results show that NISQ devices do not achieve fair sampling yet. We also observe differences in the software stack with a  particular focus on compilation techniques that illustrate what work will still need to be done to achieve a seamless integration of frontend (i.e. quantum circuit description) and backend compilation. 

\end{abstract}

\section{Introduction}

While finding \emph{any} optimal solution is usually the goal when solving optimization problems, sampling fairly from all optimal solutions is an essential component of many real-life optimization applications, such as satisfiability-based probabilistic membership filters \cite{azinovic2017assessment}, detecting equally likely fluid flow outcomes in subsurface modeling problems \cite{harp2008aquifer,o2018approach}, and generally in engineering/physics contexts where the objective function does not explicitly encode all design goals. 

Fair sampling from all optimum solutions has been proposed \cite{golden2021qaoa} as a benchmark problem for Noisy Intermediate Scale Quantum (NISQ) devices, inspired by theoretical work on fairness in quantum annealing \cite{K_nz_2019}. In this paper, we thoroughly examine the state of fair sampling on a wide range of NISQ platforms by considering both the probability that the device finds an optimum solution (which we call ground-state probability) and the fairness metric from \cite{golden2021qaoa} based on a statistical test on how many runs are needed to reject the hypothesis that device is actually a fair sampler. The NISQ platforms that we consider are seven backends from IBM~Q ranging in Quantum Volume from 8 to 128, the Aspen-9 Rigetti device, an IonQ device and three different D-Wave quantum annealing devices. We accessed IonQ, Rigetti, and two of the D-Wave devices through the Amazon Braket cloud service and had direct access to IBM~Q and the third D-Wave device.

We choose five standard small optimization problems (from \cite{K_nz_2019} and \cite{golden2021qaoa}) that are paradigmatic examples for fair sampling formulated as Ising problems, which can be solved by the Grover Mixer Quantum Alternating Operator Ansatz (GM-QAOA) \cite{baertschi2020grover} for general gate-level quantum devices (i.e., IBM~Q, Rigetti, and IonQ) and by the standard annealing algorithm for Ising problems on D-Wave quantum annealers. The GM-QAOA algorithm theoretically guarantees fair sampling, while the quantum annealing algorithm only samples fairly at very short annealing times at the cost of lower ground state probabilities (see theoretical analysis in \cite{K_nz_2019}). We present this background information in more detail in Section~\ref{sec:background}. Our results give a good glimpse of how far practice and theory still diverge, due to hardware and software limits.

If we had error-corrected quantum computing, we could just code up our five example problems in a standard quantum circuit language, such as IBM's QISKIT, Amazon Braket, or Rigetti's Quil and compile it into a standard intermediate representation (such as Microsoft's QIR or IBM's QASM) and then let a backend compiler turn it into machine code for the individual vendor-specific backends. The reality of 2021, however, is that such compilation can be done, but it results in lengthy circuits whose execution essentially returns just noise on current NISQ platforms. Until we have more advanced optimization passes for quantum compilers, we find that hand-optimization is a necessity. We describe our resulting circuits in Section \ref{sec:circuits}.
While we will go into some detail regarding the compilation choices for IBM~Q and D-Wave in separate Sections~\ref{sec:results_ibm_compilation} and~\ref{sec:results_dwave}, we report our comparative results along the two axes of probability of returning ground state and fairness in Section \ref{sec:results}. Our main findings are:

\begin{table*}[ht!]
\centering
	\begin{tabular}{|l|l|l|lcr|}
    	\hline
	\textbf{Problem}
	& \textbf{Ising Hamiltonian} $H_C$ 
	& \textbf{Ground States} 
	& \multicolumn{3}{l|}{\textbf{1-round GM-QAOA Optimum}}	\\
	with Diagram
	& to be minimized
	&  $\ket{(\lqbit_0 := \,\uparrow) \ldots \lqbit_{n-1}}$ 
	& $(\beta,\gamma) \div \pi$
	& $\bgstateT H_C \bgstate$
	& GSP \\
        \hline
	(a) \hspace*{-0.2cm} 
	\begin{minipage}{2.75cm}
	\includegraphics[width=2.75cm, height=2cm]{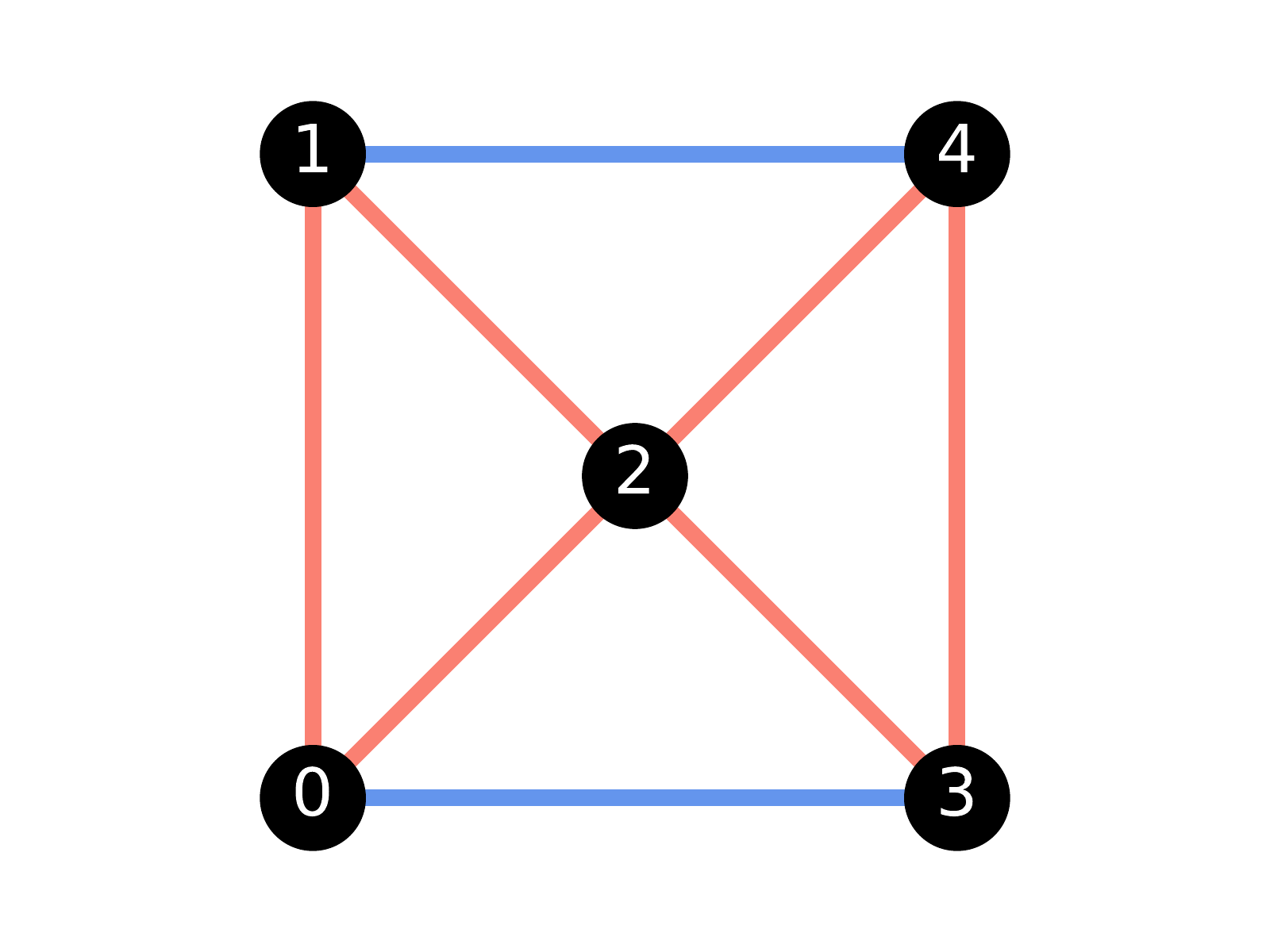}
	\end{minipage}
	& $\begin{aligned} 
		& -[ Z_0(Z_1+Z_2-Z_3) \\
		& \qquad+Z_1(Z_2-Z_4) \\
		& \qquad+Z_2(Z_3+Z_4)+Z_3Z_4 ] 
	\end{aligned}$
	& $\begin{aligned}
		& \ket{\uparrow \uparrow \uparrow \uparrow \uparrow}, \\
		& \ket{\uparrow \uparrow \uparrow \downarrow \downarrow}, \\
		& \ket{\uparrow \uparrow \downarrow \downarrow \downarrow}
	\end{aligned}$ 
	& $\left(\frac{1}{2},\frac{11}{12}\right)$
	& -2.682 out of -4
	& 0.498	\\
	(b)  \hspace*{-0.2cm}
	\begin{minipage}{2.75cm}
	\includegraphics[width=2.75cm, height=2cm]{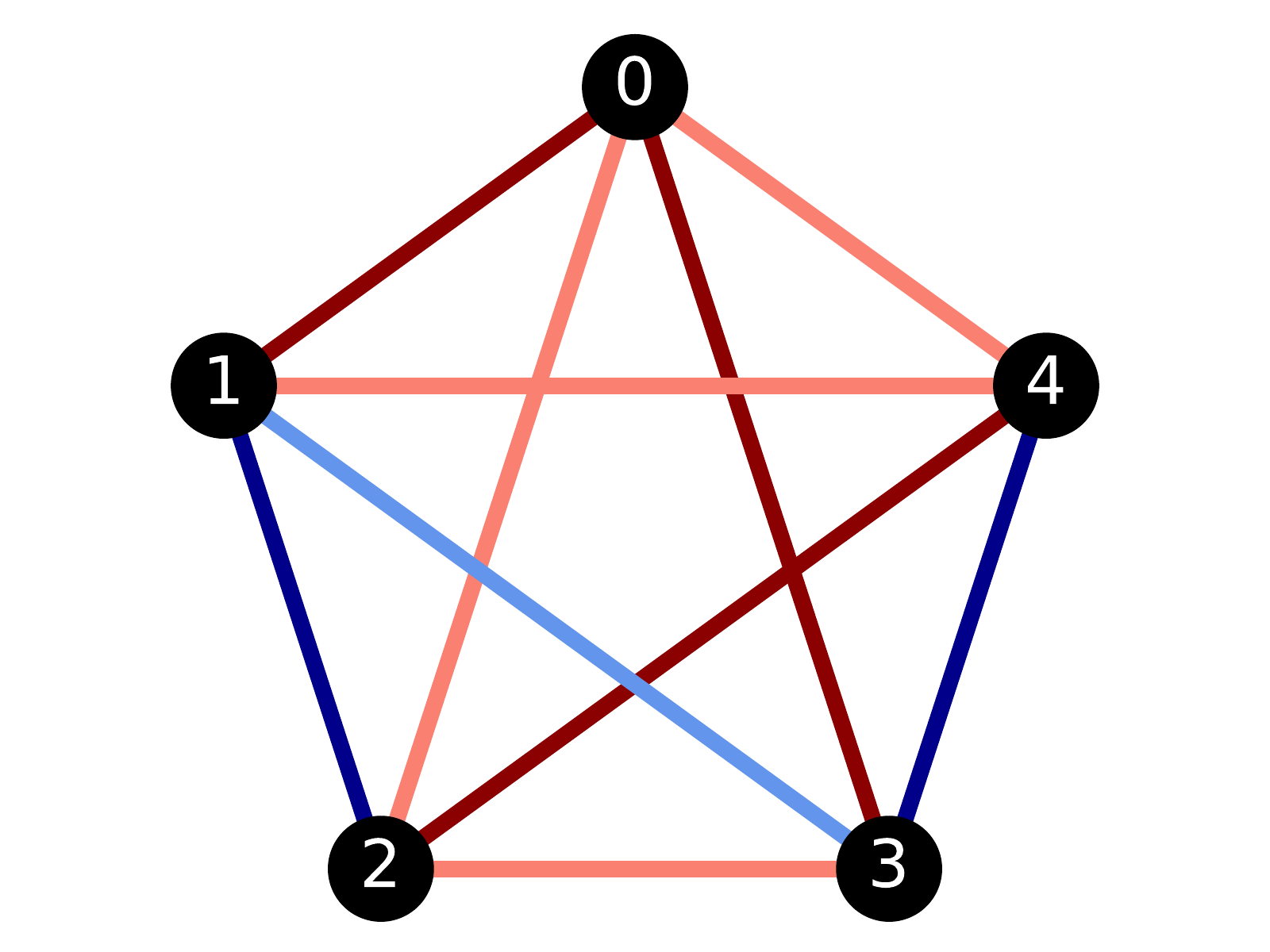}
	\end{minipage}
	& $\begin{aligned} 
		& -[ Z_0 (2 Z_1 + Z_2 + 2 Z_3 + Z_4)	\\
		& \quad+Z_1 (-2 Z_2 - Z_3 + Z_4	\\
		& \quad+Z_2(Z_3+2Z_4) - 2 Z_3 Z_4 ] 
	\end{aligned}$
        & $\begin{aligned} 
		& \ket{\uparrow \uparrow \uparrow \uparrow \uparrow},\ \ket{\uparrow \uparrow \uparrow \downarrow \uparrow},	\\
		& \ket{\uparrow \uparrow \downarrow \uparrow \downarrow},\ \ket{\uparrow \uparrow \downarrow \downarrow \uparrow},	\\
		& \ket{\uparrow \downarrow \uparrow \uparrow \uparrow},\ \ket{\uparrow \downarrow \uparrow \uparrow \downarrow}
	\end{aligned}$  
	& $\left(\frac{11}{15},\frac{17}{60}\right)$	
	& -4.228 out of -5	
	& 0.846	\\
        (c)  \hspace*{-0.2cm}
        \begin{minipage}{2.75cm}
        \includegraphics[width=2.75cm, height= 2cm]{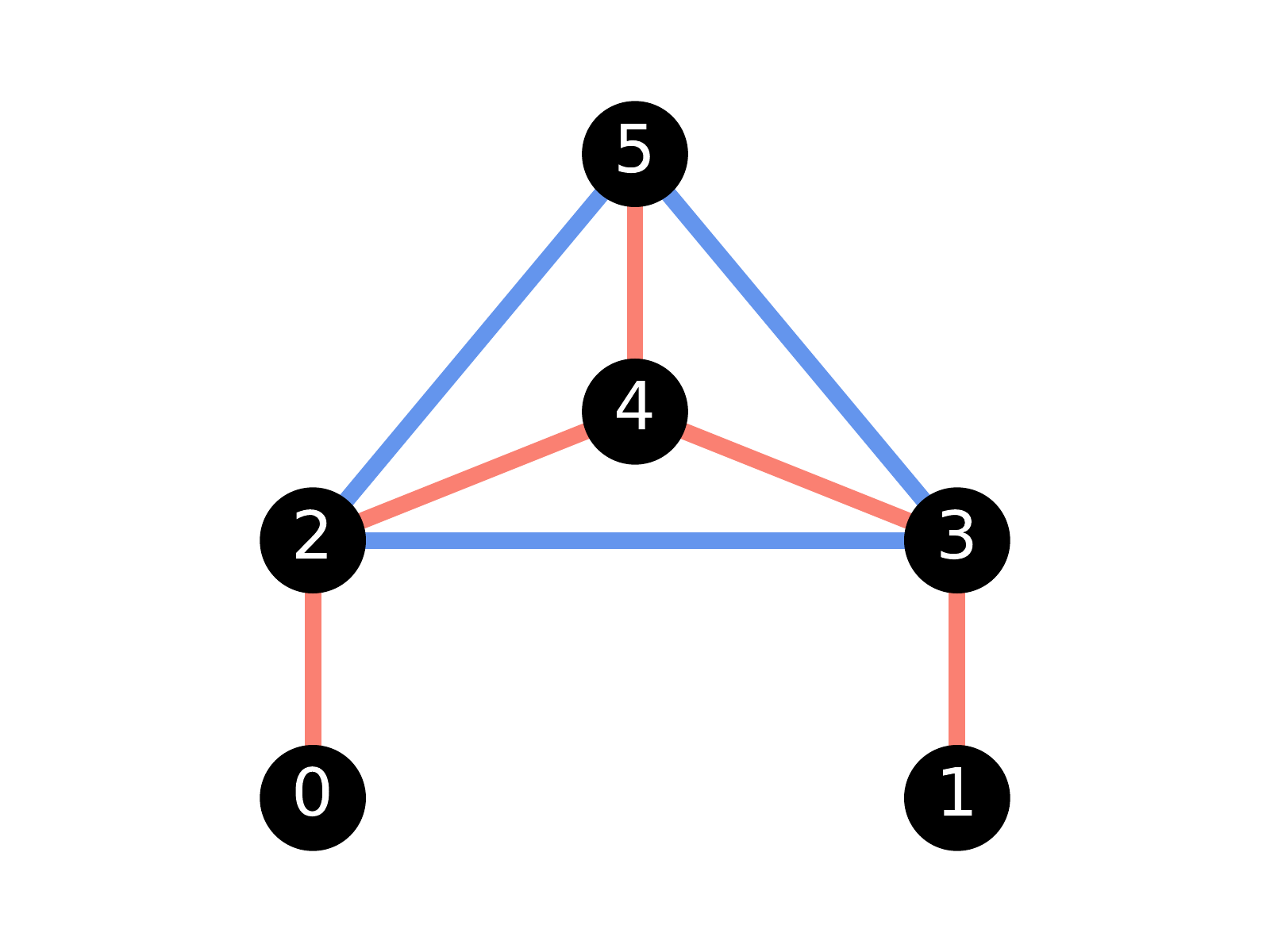}
        \end{minipage}
	& $\begin{aligned}
		& -[ Z_0 Z_2 + Z_1 Z_3	\\
		& \qquad + Z_2(-Z_3+Z_4-Z_5)	\\
		& \qquad+Z_3(Z_4-Z_5)+Z_4Z_5 ] 
	\end{aligned}$
	& $\begin{aligned}
		& \ket{\uparrow \uparrow \uparrow \uparrow \uparrow \downarrow},	\\
		& \ket{\uparrow \downarrow \uparrow \downarrow \uparrow \uparrow},	\\
		& \ket{\uparrow \downarrow \uparrow \downarrow \downarrow \downarrow}
	\end{aligned}$  
	& $\left(\frac{23}{60},-\frac{1}{15}\right)$	
	& -1.563 out of -4	
	& 0.215	\\
	(d)  \hspace*{-0.2cm}
        \begin{minipage}{2.75cm}
        \includegraphics[width=2.75cm, height= 2cm]{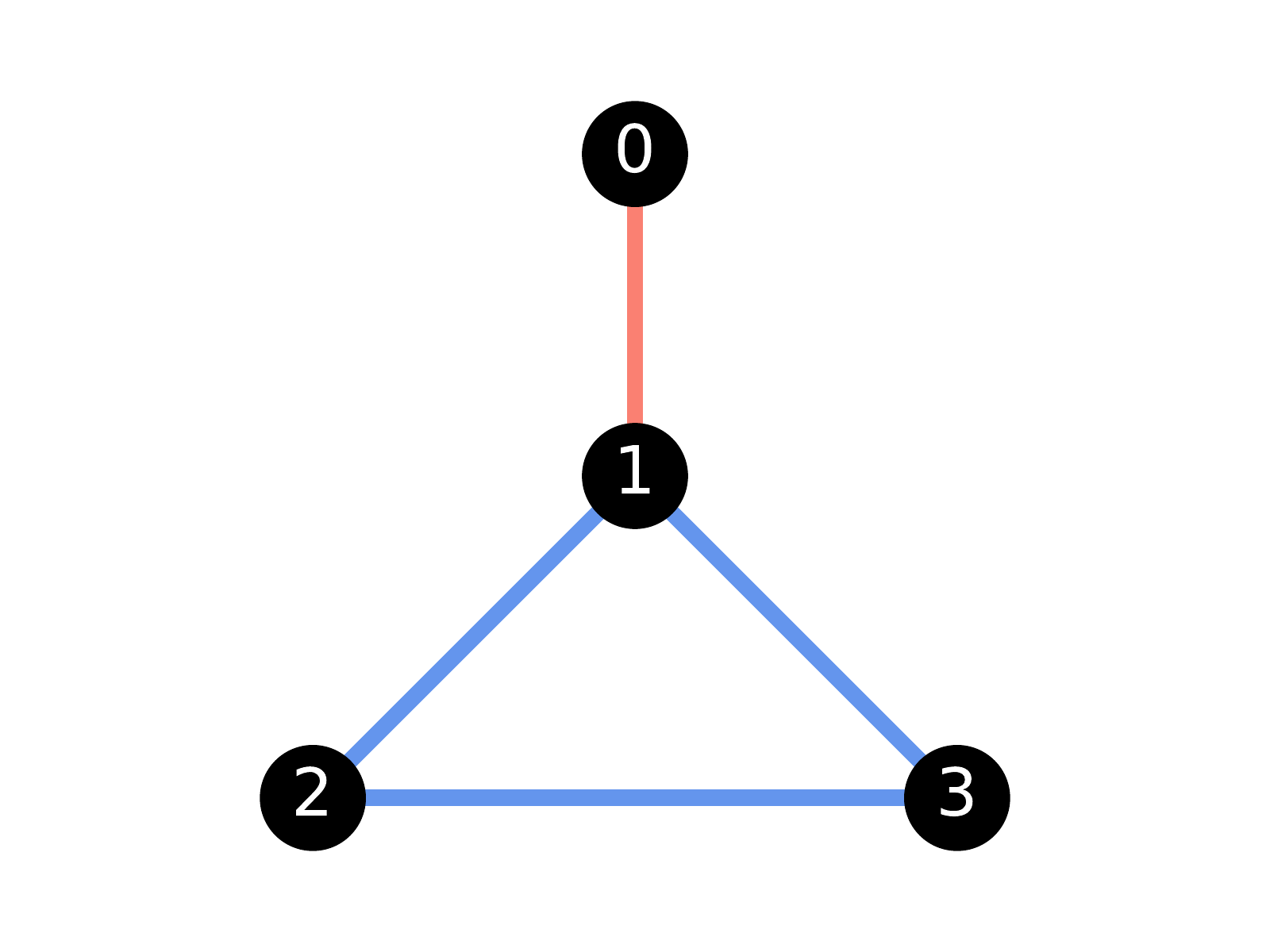}
        \end{minipage}
	& $\begin{aligned} 
		& -[ Z_0 Z_1	\\
		& \qquad+Z_1(-Z_2-Z_3) - Z_2Z_3 ]
	\end{aligned}$
	& $\begin{aligned}
		& \ket{\uparrow \uparrow \uparrow \downarrow},	\\
		& \ket{\uparrow \uparrow \downarrow \uparrow},	\\
		& \ket{\uparrow \uparrow \downarrow \downarrow}
	\end{aligned}$  
	& $\left(\frac{5}{12},-\frac{1}{10}\right)$	
	& -1.319 out of -2	
	& 0.702	\\
	(e)  \hspace*{-0.2cm}
        \begin{minipage}{2.75cm}
        \includegraphics[width=2.75cm, height= 2cm]{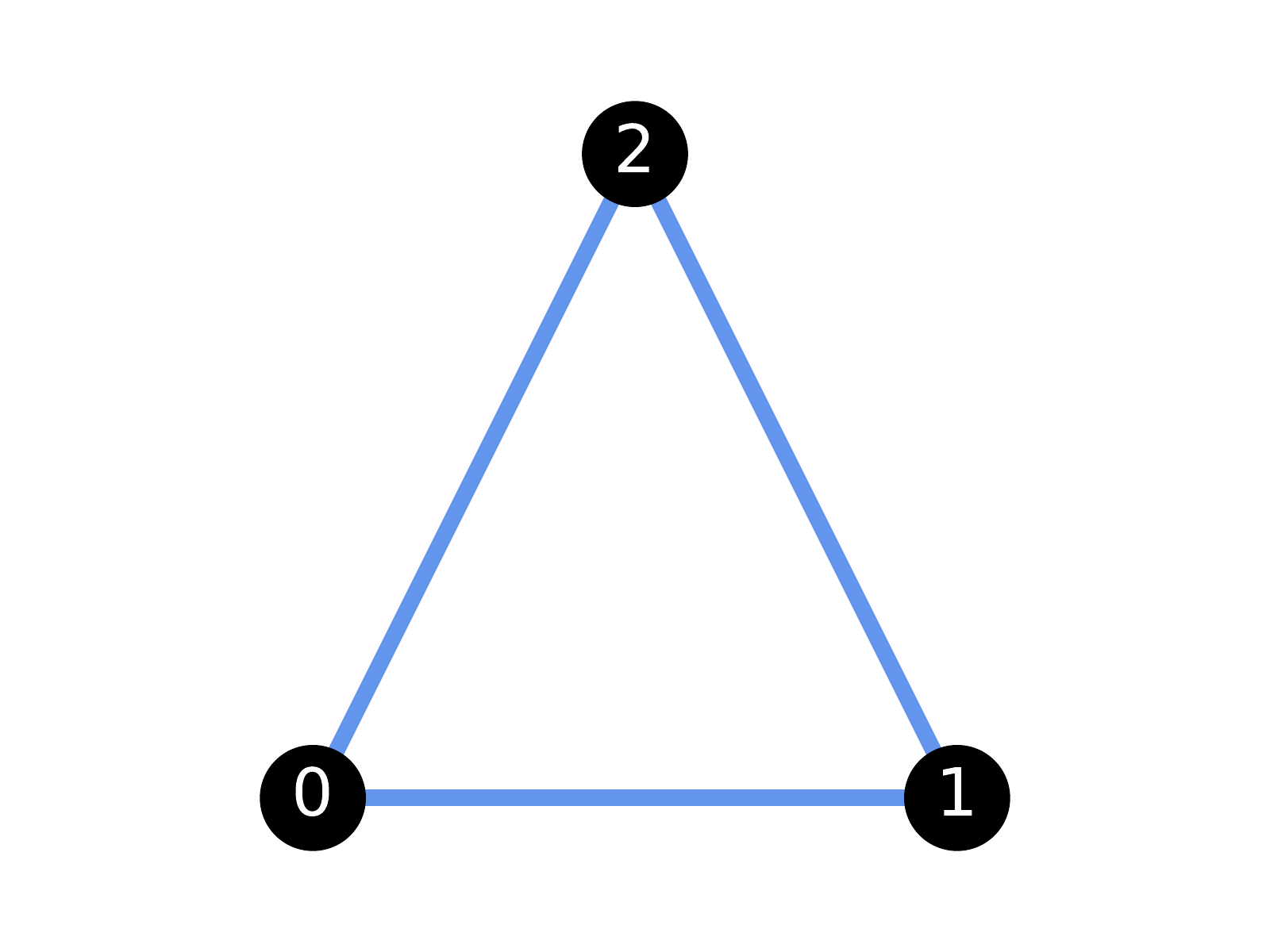}
        \end{minipage}
	& $-[ Z_0(-Z_1-Z_2)-Z_1Z_2 ] $
	& $\begin{aligned}
		& \ket{\uparrow \uparrow \downarrow},	\\
		& \ket{\uparrow \downarrow \uparrow},	\\
		& \ket{\uparrow \downarrow \downarrow}
	\end{aligned}$  
	& $\left(\frac{23}{60},-\frac{3}{5}\right)$	
	& -0.999 out of -1	
	& 1.000		\\
	\hline
    	\end{tabular}
	\caption{Ising models with degenerate ground states to be studied on NISQ hardware.
		\textbf{Problems} (a)--(d) are from~\cite{K_nz_2019}. 
		The dark red edges indicate a ferromagnetic $J_{ij} = +2$ coupling and the light red edges a ferromagnetic $J_{ij}=+1$ coupling;
		the light blue edges represent an antiferromagnetic $J_{ij} = -1$ coupling and the dark blue edges an antiferromagnetic 
		$J_{ij}=-2$ coupling.
		All \textbf{Ising Hamiltonians} $H_C = -\sum J_{ij} Z_i Z_j$ have only quadratic terms and no linear terms.
		Note that a symmetry-breaking fixed setting of the logical qubit $\lqbit_0 := \, \uparrow$ results in a Hamiltonian 
		$H_C'$ on logical qubits $\lqbit_1,\ldots,\lqbit_{n-1}$ (without $\lqbit_0$) with some linear terms.
		Only \textbf{Ground States} with $\lqbit_0=\,\uparrow$ are listed as all models are symmetric 
		under simultaneous $\uparrow/\downarrow$-flips.
		The right column shows the \textbf{Optimum Energy} $\bgstateT H_C \bgstate$ of a 1-round Grover Mixer QAOA found with a fine grid search 
		for angles $(\beta,\gamma)$ with a grid resolution of $\frac{1}{60}\cdot \pi$ and the corresponding ground state probabilities.
	}
	\label{tab:models}
\end{table*}

\subsubsection{IBM~Q devices achieve the highest fairness with IonQ a close second} In all examples, the higher quantum volume IBM~Q backends achieve the best fairness values with IonQ matching the IBM performance in 4 out of the 5 examples. Rigetti matches IBM's performance in 2 of the examples, but falls behind by 50 \% in the remaining examples.

\subsubsection{D-Wave finds ground states most reliably} D-Wave typically ends up in ground states more then 99 \% of the runs, where as the gate-level platforms struggle from around 8 \% for the hardest example to around 85 \% for simpler cases. The primary explanation for their poor performance is that we only allowed for a 1-round Grover Mixer QAOA; the gate-level devices achieve usually 10 to 40 \% less ground state probability than noise-free simulations would predict, so we would not expect 100 \% even in a noise-free environment. Allowing for more rounds and thus longer circuits would increase the theoretical optimum achievable, but in practice would lead to a further decrease in ground state probabilities as the noise just dominates the algorithmic improvement of additional rounds.
\subsubsection{IonQ finds ground states most reliably among gate-level machines with IBM a close second} IonQ performs best with respect to finding ground states compared to Rigetti and IBM in 3 of the examples. IBM wins in the other 2 examples with Rigetti's performance matching IBM's in one of them. 
\subsubsection{Rigetti's circuits have the highest aggregate error rate} Rigetti's relatively poor performance can be attributed to its high aggregate error rates. Using vendor provided qubit-level error rates, we calculate the aggregate error of each circuit from the examples. Rigetti's error rates are usually higher than 80 \%, IonQ's is around 50 \%, and IBM~Q error rates span a wide range, largely corresponding to the different device generation and achieved Quantum volumes.

\begin{table*}[t!]
	\centering
	\begin{tabular}{|p{2.35cm}|l|l|l|l|l|l|}
		\hline
		Name
		& \textbf{LNN}
		& \textbf{5T}
		& \textbf{5P}
		& \textbf{6A}
		& \textbf{7H}
		& \textbf{Clique}	\\
		\hline
		\begin{minipage}{2.2cm}
		Hardware\\ Sub-Topology
		\end{minipage}
		& \begin{minipage}{2cm}
{\scriptsize
\begin{verbatim}
pq0-pq1-pq2




\end{verbatim}}
		\end{minipage}
		& \begin{minipage}{2cm}
{\scriptsize
\begin{verbatim}
pq0-pq1-pq4
     |
    pq2
     |
    pq3
\end{verbatim}}
		\end{minipage}
		& \begin{minipage}{1.5cm}
{\scriptsize
\begin{verbatim}
pq0-pq3
 |   |
pq1-pq2
 |
pq4
\end{verbatim}}
		\end{minipage}
		& \begin{minipage}{1.5cm}
{\scriptsize
\begin{verbatim}
pq0-pq3
 |   |
pq1-pq2
 |   |
pq4 pq5
\end{verbatim}}
		\end{minipage}
		& \begin{minipage}{2cm}
{\scriptsize
\begin{verbatim}
pq0     pq4
 |       |
pq1-pq2-pq3
 |       |
pq5     pq6
\end{verbatim}}
		\end{minipage}
		& \begin{minipage}{2cm}
{\scriptsize
\begin{verbatim}
all-to-all




\end{verbatim}}
		\end{minipage}\\
		\hline
		Devices
		& All IBMQ, Rigetti
		& All IBMQ, Rigetti
		& \multicolumn{2}{l|}{IBMQ Melbourne, Rigetti}
		& All IBMQ
		& IonQ	\\
		\hline
	\end{tabular}
	\caption{Hardware topologies we compile our circuits to. The topologies correspond to graphs of 
		physical qubits $\pqbit_0, \ldots, \pqbit_6$ available as sub-topologies on IBM, Rigetti, and IonQ devices: 
		LNN (Linear Nearest Neighbor), 5T, 5P, 6A, 7H (labeled by shape) and Clique (full connectivity). 
		We compile the smaller problems (d)--(e) to LNN and the larger problems (a)--(c) to all topologies they fit on, using only the available gates of the corresponding device(s).
		For D-Wave embedding, see Section~\ref{sec:methods_DWave}.
	}
	\label{table:topologies}
\end{table*}

\subsubsection*{Details} The detailed scatter plots in Section \ref{sec:results} provide more insights. The Pareto front of backends in a fairness vs. ground-state probability plot is dominated by IBM backends with a few notable IonQ experiments interspersed. If we include D-Wave, D-Wave dominates the Pareto front on the ground state probability but only at a typically low fairness level. Recall that a data point is on the Pareto front if no other point exists that is better in both fairness and ground state probability.

We present more detailed results on the IBM~Q compiler environment in connection with our examples in Section \ref{sec:results_ibm_compilation}. The compiler passes and flags that IBM~Q QISKIT offers produce at times unpredictable results. We look at the aggregate circuit error that we can compute without actually running the circuit. We find that selecting noise adaptive compilation option reduces errors often, albeit increasing it in other cases. A similar effect can be observed when manually selecting the circuit topology preserving initial qubit layout for the circuits. Since the changes in aggregate error rates for these circuits can achieve reductions from 80 \% error to less than 30 \%, our results suggest that users should always test a large number of compiler options. We also look at QISKIT's four different optimization levels for compilation (0,1,2,3,). These levels generally reduce aggregate error by a few percentage points as we increase the optimization level; however, optimization level 1 shows erratic behavior at times tripling error rates over optimization level 0, but at other times actually beating even level 3 by 10\%.

Section \ref{sec:results_dwave} presents more detailed results on studying the effect of varying the annealing time parameter on D-Wave from 1 to 300 microseconds. In three of our examples, we see the theoretically predicted increase in ground state probability between 1 and 10 microseconds at a significant level. Overall we conjecture that even the minimum value of 1 microsecond for annealing time is already too large to still allow for fairness, thus real-life D-Wave is good at finding optimum solutions, but does not sample fairly among them.

Overall, our results characterize the state of fair sampling on a good subset of all available NISQ devices. While IBM and IonQ outperform Rigetti and D-Wave in our fairness metric, we want to be clear that none of the tested NISQ platforms are very fair in the first place. Much work remains to be done in quantum hardware design to truly achieve fair sampling.

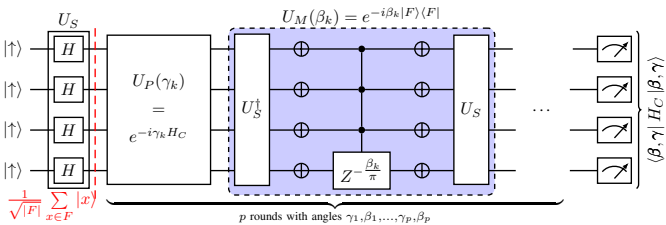
\begin{figure}[b!]
	\centering
	\begin{adjustbox}{width=\linewidth}
	\newcommand{\zgate}[1]{\gate{Z^{-\tfrac{\beta_{#1}}{\pi}}}}
	\hspace*{-0.5cm}
	\begin{quantikz}[row sep={24pt,between origins},execute at end picture={
				\node[xshift=-10pt, yshift=-20pt] at (\tikzcdmatrixname-4-2) {
					\smash{\textcolor{red}{$\frac{1}{\sqrt{|F|}}\sum\limits_{x\in F} \ket{x}$}}
				};	
				\node[xshift=20pt, yshift=-24pt] at (\tikzcdmatrixname-4-5) {
					$\underbrace{\hspace*{270pt}}_{p\text{ rounds with angles } \gamma_1,\beta_1,\ldots,\gamma_p,\beta_p}$
				};
			}]
		\lstick{$\ket{\uparrow}$}	
		& \gate{H}
		\gategroup[4,steps=1,style={inner xsep=-0pt,inner ysep=-3pt,yshift=1pt}]{$U_S$}\slice{}
		& \gate[4]{\phantom{I_I}\smash{\begin{matrix} U_P(\gamma_k) \\[1ex] = \\[1ex] e^{-i\gamma_k H_C} \end{matrix}}\phantom{I_I}}
		& \gate[4]{U_S^{\smash{\dagger}}}
		\gategroup[4,steps=5,style={dashed,rounded corners,fill=blue!20,inner sep=0pt},background]{$U_M(\beta_k) = e^{-i\beta_k \ket{F}\bra{F}}$}
		& \targ{}
		& \ctrl{1}
		& \targ{}
		& \gate[4]{U_S}
		& \qw\midstick[4,brackets=none]{\ldots}
		& \meter{}\rstick[4]{\rotatebox{90}{$\bgstateT H_C \bgstate$}}
		\\	
		\lstick{$\ket{\uparrow}$}	& \gate{H}	&&	& \targ{} 	& \ctrl{1}	& \targ{} 	&	&\qw	& \meter{}		\\	
		\lstick{$\ket{\uparrow}$}	& \gate{H}	&&	& \targ{}	& \ctrl{1}	& \targ{}	&	&\qw	& \meter{}		\\	
		\lstick{$\ket{\uparrow}$}	& \gate{H}	&&	& \targ{}	& \zgate{k}	& \targ{}	&	&\qw	& \meter{}	
	\end{quantikz}
	\end{adjustbox}
	\caption{GM-QAOA: State preparation $U_S$ gives equal superposition of all feasible states $\ket{F} = \smash{|F|^{-1/2}\sum_{x\in F} \ket{x}}$.
		$U_S$ and $U_S^{\dagger}$ are used to implement the Mixer $U_M(\beta) = e^{-i\beta \ket{F}\bra{F}}$.
		For unconstrained problems, we have 
		$U_S\ket{\uparrow^n} = H^{\otimes n}\ket{\uparrow^n} = \ket{\rightarrow^n}$ 
		and $U_M(\beta) = e^{-i\beta \ket{\rightarrow^n}\bra{\rightarrow^n}}$. 
		%
		The multi-control-$Z^{-\beta/\pi}$ gate is fully symmetric, thus we may swap controls and target.		
	}
	\label{fig:gmqaoa}
\end{figure}

\section{Background: Setting the Stage}
\label{sec:background}

In this section we describe previous studies of fair sampling in the context of quantum computing. 
A note on nomenclature: since the optimization problems we study can all be phrased in terms of Ising models, we adopt the more physics-oriented language and refer to `optimal solutions' as ground states. 
Problems with multiple optimal solutions are thus said to have degenerate ground states. 
We follow the tradition of analyzing and assessing fair sampling in the quantum computing context as defined for quantum annealing \cite{K_nz_2019} and general gate-level quantum computing \cite{golden2021qaoa}. Taking the paradigmatic examples suggested by these earlier works, we define five standard examples of optimization problems, whose graph representation is shown in the first column of Table \ref{tab:models} with the formal objective function expressed as Ising Hamiltonians in the second column, and the resulting optimal solutions or ground states (expressed in an $\uparrow,\downarrow$ basis) in the third column. 
Problems (a)-(d) all exhibit theoretically understood biased sampling with quantum annealing, so they are a somewhat natural non-trivial challenge for fair sampling algorithms. We added Problem (e) as an even smaller test case for our studies. Since all problems are less than 7 qubits, it appears plausible that they could be solved with reasonably well on current NISQ hardware.

By way of giving background, basic quantum annealing \cite{kadowaki1998quantum} is known to not always sample degenerate ground states fairly from a theoretical \cite{matsuda2009ground} and experimental \cite{mandra2017exponentially} perspective. Several approaches to improve fairness of quantum annealing have been made \cite{sieberer2018programmable,K_nz_2019,yamamoto2020fair,kumar2020achieving}. An additional suggestion is to add to the transverse field driver Hamiltonian $\sum X_i$ all higher-order Pauli-$X$ terms,
resulting in $\sum X_i + \sum X_i X_j + \sum X_i X_j X_k + \ldots = \ket{\rightarrow^n}\bra{\rightarrow^n}$~\cite{matsuda2009ground}. The study of limited higher-order driver Hamiltonians (i.e., without going over the complete sum), led K\"{o}nz et al \cite{K_nz_2019} to introduce the paradigmatic Problems (a)-(d). 
As Ising problems consisting of sums of quadratic $ZZ$ terms, the D-Wave quantum annealer takes these as native inputs. For gate-level NISQ devices, we need an appropriate algorithm to solve the problems. The most natural candidate for this is the Grover Mixer QAOA (GM-QAOA) algorithm \cite{baertschi2020grover}, which in theory samples fairly.

\begin{figure*}[ht!]
	\centering
	\begin{adjustbox}{width=\textwidth}
	\hspace*{-0.25in}
	\newcommand{\zgate}[1]{\gate{Z^{-\tfrac{\beta_{#1}}{\pi}}}}
	\newcommand{\rzgate}[1]{\gate{R_z(#1\gamma)}}
	\newcommand{\sxgate}{\gate{\sqrt{X}}}
	\begin{quantikz}[row sep={24pt,between origins},execute at end picture={
				\draw[semithick,gray,-,rounded corners]	($(\tikzcdmatrixname-2-1)-(60pt,0pt)$) to [bend right] ($(\tikzcdmatrixname-3-1)-(60pt,0pt)$);
				\draw[semithick,gray,-,rounded corners]	($(\tikzcdmatrixname-3-1)-(60pt,0pt)$) to [bend right] ($(\tikzcdmatrixname-4-1)-(60pt,0pt)$);
				\draw[semithick,gray,-,rounded corners]	($(\tikzcdmatrixname-4-1)-(60pt,0pt)$) to [bend right] ($(\tikzcdmatrixname-5-1)-(60pt,0pt)$);
				\draw[semithick,gray,-,rounded corners]	($(\tikzcdmatrixname-2-1)-(60pt,0pt)$) to [bend right] ($(\tikzcdmatrixname-5-1)-(60pt,0pt)$);
				\draw[semithick,gray,-,rounded corners]	($(\tikzcdmatrixname-3-1)-(60pt,0pt)$) to [bend right] ($(\tikzcdmatrixname-6-1)-(60pt,0pt)$);
				\draw[semithick,gray,-,rounded corners]	($(\tikzcdmatrixname-4-1)-(60pt,0pt)$) to [bend right] ($(\tikzcdmatrixname-7-1)-(60pt,0pt)$);
				\draw[thick,-,rounded corners]	(\tikzcdmatrixname-2-7) to [bend right] (\tikzcdmatrixname-5-7);
				\draw[thick,-,rounded corners]	(\tikzcdmatrixname-2-9) to [bend left] (\tikzcdmatrixname-5-9);
				\draw[thick,-,rounded corners]	(\tikzcdmatrixname-2-10) to [bend left] (\tikzcdmatrixname-5-10);
				\node[fit=(\tikzcdmatrixname-1-15)(\tikzcdmatrixname-5-15),draw,cross out,thick,blue] {};				
				\node[fit=(\tikzcdmatrixname-1-17)(\tikzcdmatrixname-5-17),draw,cross out,thick,blue] {};	
				\node[fit=(\tikzcdmatrixname-1-23)(\tikzcdmatrixname-1-25),draw,cross out,thick,blue] {};
				\node[fit=(\tikzcdmatrixname-2-3)(\tikzcdmatrixname-6-3),draw,cross out,thick,red] {};				
				\node[fit=(\tikzcdmatrixname-2-20)(\tikzcdmatrixname-6-20),draw,cross out,thick,red] {};				
				\node[fit=(\tikzcdmatrixname-2-22)(\tikzcdmatrixname-6-23),draw,cross out,thick] {};				
				\node[fit=(\tikzcdmatrixname-2-27)(\tikzcdmatrixname-6-27),draw,cross out,thick,red] {};
		}]
		\lstick{N/A$\colon \lqbit_0\ \ket{\uparrow}$}	
		& \qw	& \qw	& \qw	& \qw	& \qw	& \qw	& \qw	& \qw	& \qw	& \qw	& \qw	& \qw	& \qw	
		& \ctrl{4}
		& \qw
		& \ctrl{4} 
		& \qw	& \qw	& \qw	& \qw	& \qw
		& \targ{}
		& \ctrl{1}
		& \targ{}
		& \qw	& \qw
		& \qw\bra{\uparrow}\rstick[6]{\rotatebox{90}{$\bgstateT H_C \bgstate$}}
		\\	
		\lstick{$\pqbit_0\colon	\lqbit_2\ \ket{\uparrow}$}	& \sxgate	& \gate{S}	& \ctrl{1}	& \qw		& \ctrl{1}	& \ctrl{}	& \qw		& \targ{}	& \ctrl{}	& \ctrl{1}	& \qw		& \ctrl{1}	& \qw		& \qw		& \qw		& \qw		& \qw		& \qw\lqbit_5	& \gate{S^{\dagger}}	& \sxgate	& \targ{}	& \targ{}	& \ctrl{1}	& \targ{} 	& \sxgate	& \gate{S^{\dagger}}	&\meter{}	\\	
		\lstick{$\pqbit_1\colon	\lqbit_3\ \ket{\uparrow}$}	& \sxgate	& \gate{S}	& \targ{}	& \rzgate{2}	& \targ{}	& \ctrl{1}	& \qw		& \ctrl{1}	& \qw		& \targ{}	& \rzgate{2}	& \targ{}	& \ctrl{3}	& \qw		& \qw		& \qw		& \ctrl{3}	& \qw\lqbit_3	& \gate{S^{\dagger}}	& \sxgate	& \targ{}	& \targ{}	& \zgate{}	& \targ{}	& \sxgate	& \gate{S^{\dagger}}	&\meter{}	\\	
		\lstick{$\pqbit_2\colon	\lqbit_4\ \ket{\uparrow}$}	& \sxgate	& \gate{S}	& \ctrl{1}	& \qw		& \ctrl{1}	& \targ{}	& \rzgate{-2}	& \targ{}	& \qw		& \ctrl{1}	& \qw		& \ctrl{1}	& \qw		& \qw		& \qw		& \qw		& \qw		& \qw\lqbit_4	& \gate{S^{\dagger}}	& \sxgate	& \targ{}	& \targ{}	& \ctrl{-1}	& \targ{}	& \sxgate	& \gate{S^{\dagger}}	&\meter{}	\\
		\lstick{$\pqbit_3\colon	\lqbit_5\ \ket{\uparrow}$}	& \sxgate	& \gate{S}	& \targ{}	& \rzgate{-2}	& \targ{}	& \targ{}	& \rzgate{2}	& \ctrl{}	& \targ{}	& \targ{}	& \rzgate{-2}	& \targ{}	& \qw		& \targ{}	& \rzgate{-2}	& \targ{}	& \qw		& \qw\lqbit_2	& \gate{S^{\dagger}}	& \sxgate	& \targ{}	& \targ{}	& \ctrl{-1}	& \targ{}	& \sxgate	& \gate{S^{\dagger}}	&\meter{}	\\
		\lstick{$\pqbit_4\colon	\lqbit_1\ \ket{\uparrow}$}	& \sxgate	& \gate{S}	& \qw		& \qw		& \qw		& \qw		& \qw		& \qw		& \qw		& \qw		& \qw		& \qw		& \targ{}	& \qw		& \rzgate{-2}	& \qw		& \targ{}	& \qw\lqbit_1	& \gate{S^{\dagger}}	& \sxgate	& \targ{}	& \targ{}	& \ctrl{-1}	& \targ{}	& \sxgate	& \gate{S^{\dagger}}	&\meter{}	\\
		\lstick{$\pqbit_5\colon	\anc\ \ket{\uparrow}$}		& \qw		& \qw		& \qw		& \qw		& \qw		& \qw		& \qw		& \qw		& \qw		& \qw		& \qw		& \qw		& \qw		& \qw		& \qw		& \qw		& \qw		& \qw\anc	& \qw			& \qw		& \qw		& \qw		& \qw		& \qw		& \qw		& \qw			&\qw\bra{\uparrow}
	\end{quantikz}
	\end{adjustbox}
	\caption{Implementation of State Preparation $U_S$ and the Phase Separator $U_P$ for Problem (c) on a 6A-connectivity ($\pqbit_0, \ldots, \pqbit_5$, light gray) for IBM~Q:
		The logical qubit $\lqbit_0:= \ket{\uparrow}$ is fixed, effectively making all its gates and controls \textcolor{blue}{redundant}.
		\textbf{(Phase separator)}
		The weighted Ising terms $w Z_i Z_j$ in the cost Hamiltonian $H_C$ of the phase separator $U_P = e^{-i\gamma H_C}$ 
		pairwise commute and can thus be implemented individually with 2 $\CNOT$s and 1 $R_z(-2\cdot w\cdot \gamma)$. 
		IBM~Q Melbourne does not contain a $4$-clique in its topology, only a $4$-cycle, hence implementing the $4-$clique of logical qubits 
		$\lqbit_2, \lqbit_3, \lqbit_4, \lqbit_5$ needs at least one $\SWAP$. We incorporated it into the phase separator
		on logical qubits $\lqbit_2, \lqbit_5$ (drawn with curved $\CNOT$s), resulting in a permuted assignment 
		to the physical qubits at the end.
		\textbf{(State preparation)}
		Given $H\ket{\uparrow} = \ket{\rightarrow} = S\sqrt{X}\ket{\uparrow}$, the state preparation unitary and its
		inverse have been rewritten as $U_S = \smash{(S\sqrt{X})^{\otimes n-1}}$ and 
		$U_S^{\dagger} = \smash{(\sqrt{X}^{\dagger}S^{\dagger})^{\otimes n-1}} = \smash{(X\sqrt{X}S^{\dagger})^{\otimes n-1}}$. 
		The phase shifts $S$ and $S^{\dagger}$ commute through the phase separator and cancel, and the action
		before measurement in the $Z$-basis only affects phase not probability; hence they can be \textcolor{red}{removed}.
		\textbf{(Grover Mixer)}
		Schematics only, for implementation see Fig.~\ref{fig:ibm-GM}.
	}
	\label{fig:ibm-PS}
\end{figure*}
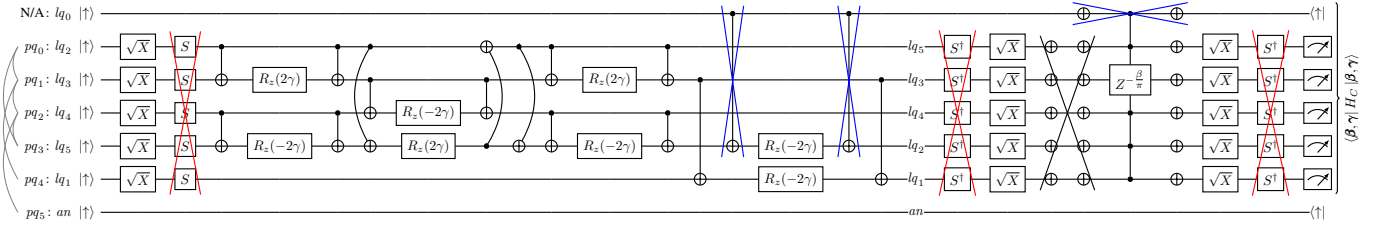

GM-QAOA is a variation of the Quantum Alternating Operator Ansatz~\cite{hadfield_qaoa}.
In its essence, for a problem instance $I$ with feasible solution states $F$ and cost Hamiltonian $H_C$ on $n$ qubits, a $p$-round QAOA prepares a parameterized state
from which it samples low-energy states with respect to $H_C$:
\begin{align}
	\bgstate := U_M(\beta_p) U_P(\gamma_p) \cdots U_M(\beta_1) U_P(\gamma_1) U_S \ket{\uparrow^n}.	\label{eq:bgstate}
\end{align}

The circuit consists of an initial \emph{state preparation} unitary operator $U_S$ that creates a superposition of all feasible solutions $F$, followed by $p$ applications of alternating parametrized \emph{phase separating} and \emph{mixing} unitaries $U_P(\gamma_k)$, $U_M(\beta_k)$ with real angle parameters $\bgamma = (\gamma_1, \ldots, \gamma_p)^T$ and $\bbeta = (\beta_1, \ldots, \beta_p)^T$, and a final measurement in the computational basis, see Fig.~\ref{fig:gmqaoa}.
The phase separating unitaries $U_P(\gamma)$ add multiplicative phase factors to the amplitudes of feasible computational basis states, with phases proportional to respective energies.  
We usually have (up to global phases) $U_P(\gamma) \cong e^{-i\gamma H_C}$, where  $H_C$ is an Ising Hamiltonian with quadratic and linear terms as found in the second column of Table \ref{tab:models} for our example problems.
The Grover mixer unitary \cite{baertschi2020grover} requires an efficient state preparation unitary $U_S$ of an equal superposition of all feasible basis states $\ket{F} = 1/\sqrt{|F|} \sum_{x\in F} \ket{x}$. $U_S$ can be used to design a mixing unitary $U_M$ resembling Grover's selective phase-shift operator~\cite{grover2005fixed,yoder2014fixed,akshay2020}, where
		$U_M(\beta) = e^{-i\beta \ket{F}\bra{F}} 
		= \Id - (1-e^{-i\beta})\ket{F}\bra{F} 
		= U_S(\Id - (1-e^{-i\beta})\ket{\uparrow}\bra{\uparrow})U_S^{\dagger}$.

Grover Mixer QAOA samples fairly: all feasible basis states begin with amplitude $1/\sqrt{|F|}$ (after state preparation with $U_S$).
The phase separating unitary $U_P(\gamma) = e^{-i\gamma H_C}$ then phases the amplitude of every basis state proportional to its energy and $\gamma$, keeping the same phase for basis states of same energy.
The mixing unitary $U_M(\beta) = \Id - (1-e^{-i\beta}) \ket{F}\bra{F}$ then deducts from all amplitudes the same weighted average of the amplitudes, $((1-e^{-i\beta})/\sqrt{|F|}) \bra{F}U_P(\gamma)U_S\ket{\uparrow}$.
Therefore, basis states with the same energy are sampled with the same amplitude. For a complete proof, see~\cite{baertschi2020grover}.

\section{Designing Circuits for NISQ Devices}
\label{sec:circuits}

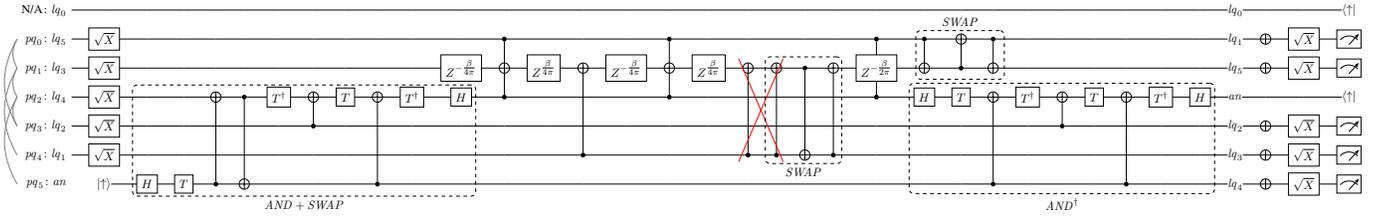
\begin{figure*}[ht!]
	\centering
	\begin{adjustbox}{width=\textwidth}
	\hspace*{-0.4in}
	\newcommand{\zgate}[2]{\gate{Z^{#1\tfrac{\beta}{#2\pi}}}}
	\newcommand{\rzgate}[1]{\gate{R_z(#1\gamma)}}
	\newcommand{\sxgate}{\gate{\sqrt{X}}}
	\newcommand{\hgate}{\gate{H}}
	\newcommand{\tgate}{\gate{T}}
	\newcommand{\tigate}{\gate{T^{\dagger}}}
	\begin{quantikz}[row sep={24pt,between origins},execute at end picture={
				\draw[semithick,gray,-]	($(\tikzcdmatrixname-2-1)-(45pt,0pt)$) to [bend right] ($(\tikzcdmatrixname-3-1)-(45pt,0pt)$);
				\draw[semithick,gray,-]	($(\tikzcdmatrixname-3-1)-(45pt,0pt)$) to [bend right] ($(\tikzcdmatrixname-4-1)-(45pt,0pt)$);
				\draw[semithick,gray,-]	($(\tikzcdmatrixname-4-1)-(45pt,0pt)$) to [bend right] ($(\tikzcdmatrixname-5-1)-(45pt,0pt)$);
				\draw[semithick,gray,-]	($(\tikzcdmatrixname-2-1)-(45pt,0pt)$) to [bend right] ($(\tikzcdmatrixname-5-1)-(45pt,0pt)$);
				\draw[semithick,gray,-]	($(\tikzcdmatrixname-3-1)-(45pt,0pt)$) to [bend right] ($(\tikzcdmatrixname-6-1)-(45pt,0pt)$);
				\draw[semithick,gray,-]	($(\tikzcdmatrixname-4-1)-(45pt,0pt)$) to [bend right] ($(\tikzcdmatrixname-7-1)-(45pt,0pt)$);
				\node[fit=(\tikzcdmatrixname-4-3)(\tikzcdmatrixname-7-12),draw,dashed,thick,rounded corners,inner xsep=12pt,inner ysep=10pt,label={below:$\mathit{AND} + \SWAP$}] {};				
				\node[fit=(\tikzcdmatrixname-3-19)(\tikzcdmatrixname-6-20),draw,cross out, red] {};
				\node[fit=(\tikzcdmatrixname-3-20)(\tikzcdmatrixname-6-22),draw,dashed,thick,rounded corners,inner sep=5pt,label={below:$\SWAP$}] {};
				\node[fit=(\tikzcdmatrixname-2-24)(\tikzcdmatrixname-3-26),draw,dashed,thick,rounded corners,inner sep=5pt,label={above:$\SWAP$}] {};
				\node[fit=(\tikzcdmatrixname-4-24)(\tikzcdmatrixname-7-32),draw,dashed,thick,rounded corners,inner xsep=8pt,inner ysep=6pt,xshift=4pt,yshift=-2pt,label={below:$\mathit{AND}^{\dagger}$}] {};				
		}]
		\lstick{N/A$\colon \lqbit_0$}	 	& \qw		& \qw		& \qw		& \qw		& \qw		& \qw		& \qw		& \qw		& \qw		& \qw		& \qw		& \qw		& \qw		& \qw		& \qw		& \qw		& \qw		& \qw		& \qw		& \qw		& \qw		& \qw		& \qw		& \qw		& \qw		& \qw		& \qw		& \qw		& \qw		& \qw		& \qw		& \qw\lqbit_0	& \qw		& \qw		&\qw\bra{\uparrow}	\\	
		\lstick{$\pqbit_0\colon	\lqbit_5$}	& \sxgate	& \qw		& \qw		& \qw		& \qw		& \qw		& \qw		& \qw		& \qw		& \qw		& \qw		& \ctrl{1}	& \qw		& \qw		& \qw		& \ctrl{1}	& \qw		& \qw		& \qw		& \qw		& \qw		& \ctrl{1}	& \ctrl{1}	& \targ{}	& \ctrl{1}	& \qw		& \qw		& \qw		& \qw		& \qw		& \qw		& \qw\lqbit_1	& \targ{}	& \sxgate	&\meter{}		\\	
		\lstick{$\pqbit_1\colon	\lqbit_3$}	& \sxgate	& \qw		& \qw		& \qw		& \qw		& \qw		& \qw		& \qw		& \qw		& \qw		& \zgate{-}{4}	& \targ{}	& \zgate{}{4}	& \targ{}	& \zgate{-}{4}	& \targ{}	& \zgate{}{4}	& \targ{}	& \targ{}	& \ctrl{3}	& \targ{}	& \zgate{-}{2}	& \targ{}	& \ctrl{-1}	& \targ{}	& \qw		& \qw		& \qw		& \qw		& \qw		& \qw		& \qw\lqbit_5	& \targ{}	& \sxgate	&\meter{}		\\	
		\lstick{$\pqbit_2\colon	\lqbit_4$}	& \sxgate	& \qw		& \qw		& \targ{}	& \ctrl{3}	& \tigate	& \targ{}	& \tgate	& \targ{}	& \tigate	& \hgate	& \ctrl{-1}	& \qw		& \qw		& \qw		& \ctrl{-1}	& \qw		& \qw		& \qw		& \qw		& \qw		& \ctrl{-1}	& \hgate	& \tgate	& \targ{}	& \tigate	& \targ{}	& \tgate	& \targ{}	& \tigate	& \hgate	& \qw\anc	& \qw		& \qw		&\qw\bra{\uparrow}	\\
		\lstick{$\pqbit_3\colon	\lqbit_2$}	& \sxgate	& \qw		& \qw		& \qw		& \qw		& \qw		& \ctrl{-1}	& \qw		& \qw		& \qw		& \qw		& \qw		& \qw		& \qw		& \qw		& \qw		& \qw		& \qw		& \qw		& \qw		& \qw		& \qw		& \qw		& \qw		& \qw		& \qw		& \ctrl{-1}	& \qw		& \qw		& \qw		& \qw		& \qw\lqbit_2	& \targ{}	& \sxgate	&\meter{}		\\
		\lstick{$\pqbit_4\colon	\lqbit_1$}	& \sxgate	& \qw		& \qw		& \qw		& \qw		& \qw		& \qw		& \qw		& \qw		& \qw		& \qw		& \qw		& \qw		& \ctrl{-3}	& \qw		& \qw		& \qw		& \ctrl{-3}	& \ctrl{-3}	& \targ{}	& \ctrl{-3}	& \qw		& \qw		& \qw		& \qw		& \qw		& \qw		& \qw		& \qw		& \qw		& \qw		& \qw\lqbit_3	& \targ{}	& \sxgate	&\meter{}		\\
		\lstick{$\pqbit_5\colon \anc$}		& \ket{\uparrow}& \hgate	& \tgate	& \ctrl{-3}	& \targ{}	& \qw		& \qw		& \qw		& \ctrl{-3}	& \qw		& \qw		& \qw		& \qw		& \qw		& \qw		& \qw		& \qw		& \qw		& \qw		& \qw		& \qw		& \qw		& \qw		& \qw		& \ctrl{-3}	& \qw		& \qw		& \qw		& \ctrl{-3}	& \qw		& \qw		& \qw\lqbit_4	& \targ{}	& \sxgate	&\meter{}
	\end{quantikz}
	\end{adjustbox}
	\caption{Grover Mixer implementation details for logical qubits $\lqbit_1,\ldots,\lqbit_5$ with fixed $\lqbit_0 := \ket{\uparrow}$ 
		on 6A-connectivity ($\pqbit_0,\ldots,\pqbit_5$, light gray) for IBM~Q:
		\textbf{(AND gates)} To reduce the $4$-control-$Z^{-\beta/\pi}$ from Fig.~\ref{fig:ibm-PS} to fewer controls, we first compute the logical $\mathit{AND}(\lqbit_2,\lqbit_4)$ 
		into an ancilla qubit $\anc$, and swap it to $\pqbit_2$. The task is now to implement a $3$-control-$Z^{-\beta/\pi}$ phase shift on central qubit $\lqbit_3$ on $\pqbit_1$
		with controls on its neighbors $\lqbit_5,\anc,\lqbit_1$ on $\pqbit_0, \pqbit_2, \pqbit_4$, before uncomputing the $\mathit{AND}$.
		\textbf{(3-control phase shift)} We use the standard decomposition into single-qubit phase shifts on $\lqbit_3$ interleaved with $\CNOT$ 
		and $\mathit{TOFFOLI}$ gates, plus a recursively smaller 2-control-$Z^{-\beta/2\pi}$ on the control qubits $\lqbit_5, \anc, \lqbit_1$. 
		As $\lqbit_1$ is not adjacent to $\lqbit_5,\anc$, we first swap it to the central qubit $\pqbit_1$. Note that due to gate cancellations,
		this $\SWAP$ only increases the $\CNOT$ count by 1.
		\textbf{(Recursion)} We recursively decompose the 2-control-$Z^{-\beta/2\pi}$ (decomposition not shown). 
		Similarly to before, we will need another $\SWAP$ (shown) to implement the final base of a 1-control-$Z^{-\beta/4\pi}$ (not shown).
		\textbf{(TOFFOLI gates)} For the two $\mathit{TOFFOLI}$s, we can avoid such a further recursion (decompositions not shown): 
		We will implement each $\mathit{TOFFOLI}$ with 2 Hadamard gates on the target conjugating a 2-control-$Z$. 
		If we interpret the second of these 2-control-$Z$ as a 2-control-$Z^{-1}$ (as $Z^{-1} = Z$), 
		the resulting recursive 1-control-$Z^{1/2}$ and 1-control-$Z^{-1/2}$ of the two $\mathit{TOFFOLI}$ implementations cancel.
		Finally, we can replace all (non-controlled) phase-shift gates by (up to global phase) equivalent $R_z$ rotations, native to IBM~Q.
	}
	\label{fig:ibm-GM}
\end{figure*}

Our general procedure for generating test circuits was to begin with a 1-round Grover Mixer QAOA algorithm 
in order to keep circuit depth low. Following~\cite{K_nz_2019}, we fix $q_0:=\,\uparrow$ as all of the models 
in Table~\ref{tab:models} are symmetric under simultaneous $\uparrow/\downarrow$ swaps on all qubits.
This reduces the problems from Ising Hamiltonians $H_C$ with only quadratic terms acting on logical qubits $\lqbit_0,\ldots,\lqbit_{n-1}$ 
to a new Hamiltonians $H_C'$ with some linear terms acting on fewer qubits $\lqbit_1,\ldots,\lqbit_{n-1}$. 
This also transforms $\lqbit_0$ into a classical control bit for the consecutive Grover Mixer, which can be removed. 
Thus we can embed Ising problems without linear terms on $n$ qubits onto circuits with only $n-1$ qubits, see Fig.~\ref{fig:ibm-PS}.
The optimum energy $\bgstateT H_C \bgstate$ achievable with a 1-round Grover Mixer QAOA was found with a fine grid search 
for angles $(\beta,\gamma)$ with a grid resolution of $\pi/60$; the values with corresponding ground state probabilities
can be found in Table~\ref{tab:models}.

We then compiled the circuits to match connectivity and gates available for each tested device individually. 
The available hardware topologies are listed in Table~\ref{table:topologies}. For IonQ and Rigetti through Braket,
there is no available compilation to the available 1- and 2-qubit gates. Compared to IBM's \verb|qiskit| compiler tools, we 
found that manual compilation reduced circuit depth by roughly a factor of two.

Hence we used compilation by hand down to the available 1- and 2-qubit gates for each individual device.
To reduce circuit depth as much as possible, we made use of three techniques: 
(i) use the available 1- and 2-qubit gates wherever possible, 
(ii) when using $\SWAP$ gates, classically track the permuted assignment 
of logical to physical qubits instead of restoring the original assignment in the circuit, 
(iii) use ancilla qubits whenever they help bring down the circuit depth of the Grover Mixer implementation.
We demonstrate our techniques in detail for the circuit compilation of Problem (c) on topology \textbf{6A}
for IBM~Q using gates $X, \sqrt{X}, \CNOT, R_z(\theta)$ in Figures~\ref{fig:ibm-PS} and~\ref{fig:ibm-GM}.
We provide accompanying interactive Quirk circuits for these figures
\href{https://algassert.com/quirk#circuit=
(state preparation \& phase separator)
and 
\href{https://algassert.com/quirk#circuit=
(multi-control-$Z^{-t}$ phase shift unitary matrix illustrated by 
\href{https://github.com/Strilanc/Quirk/wiki/How-to-use-Quirk#view-the-unitary-matrix-of-a-circuit-via-the-state-channel-duality}{Quirk's amplitude display} via state channel duality).

\section{Methodology / Experimental design}
\label{sec:methods}

\begin{figure*}[t]
    \includegraphics[trim = 0 0 0 0, clip,draft=false, height=0.3\textwidth]{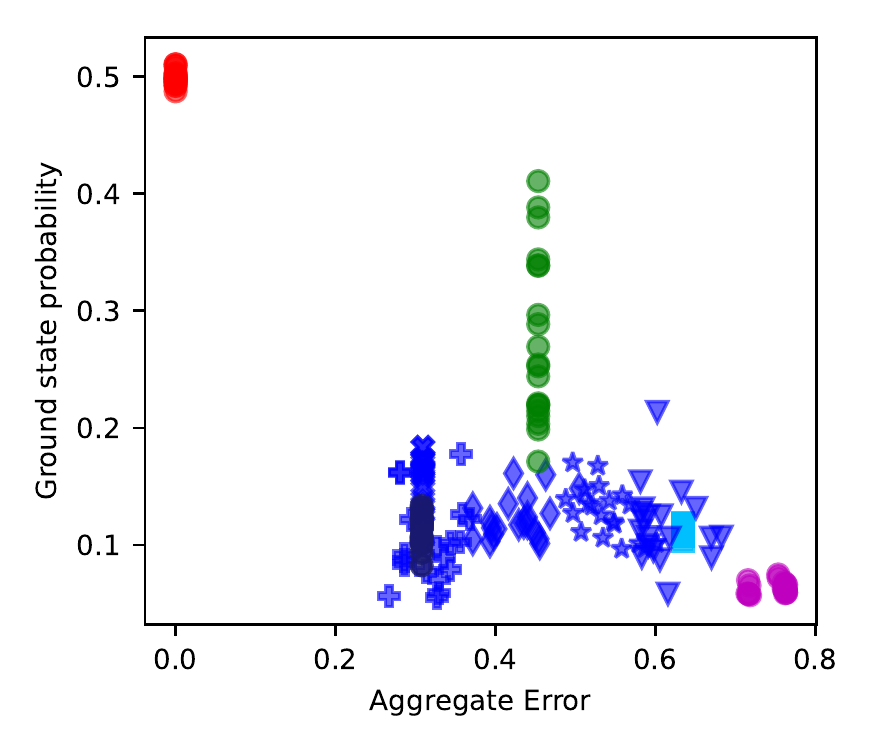}\hfill%
    \includegraphics[trim = 20 0 0 0, clip,draft=false, height=0.3\textwidth]{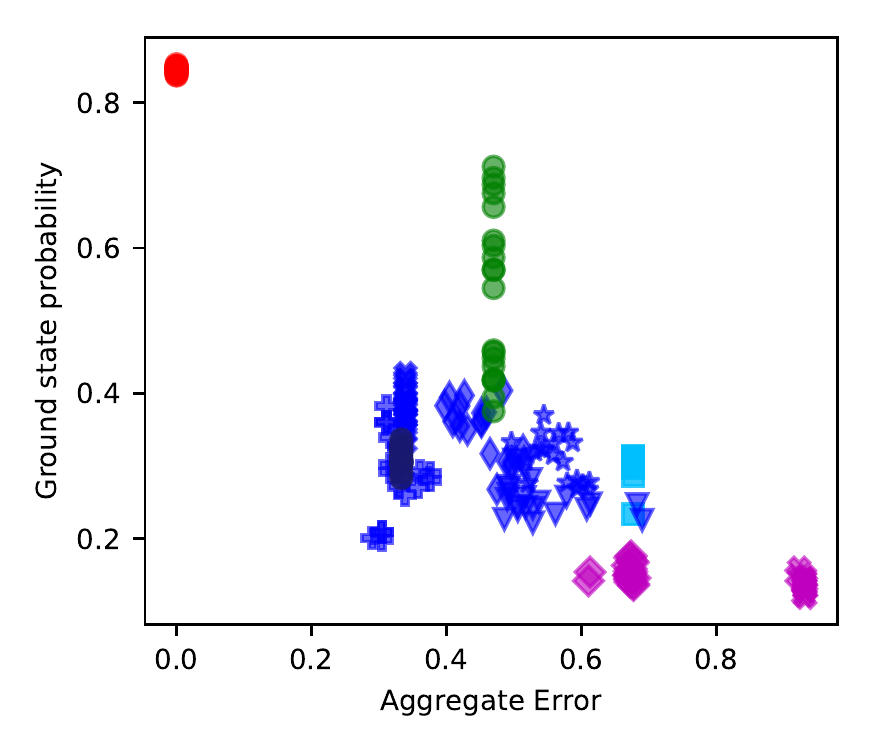}\hfill%
    \includegraphics[trim = 20 0 0 0, clip,draft=false,height=0.3\textwidth]{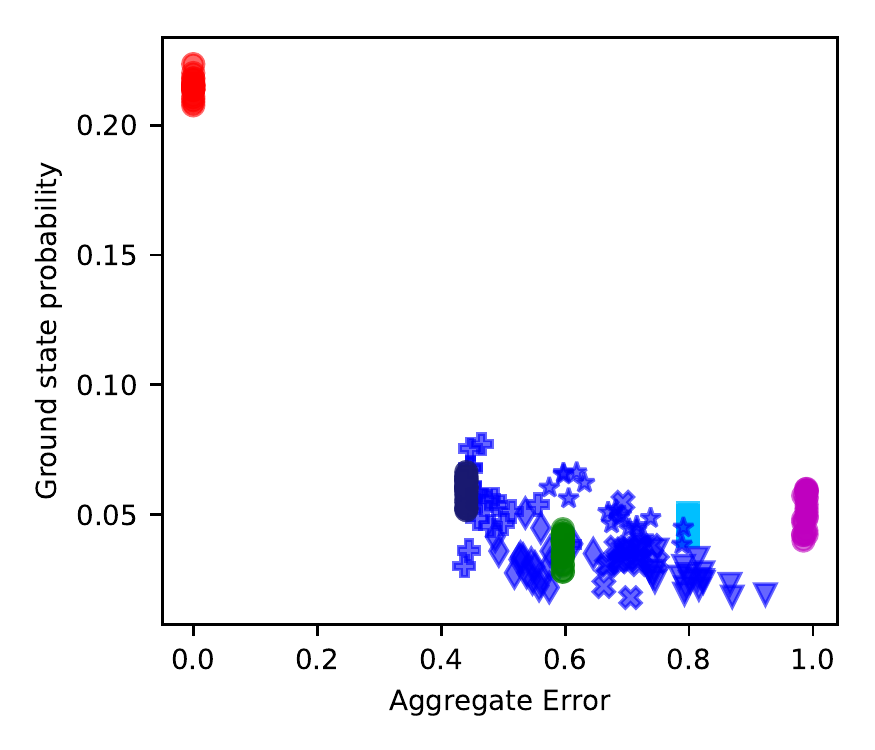}\\[-3ex]
    \hfill \includegraphics[trim = -46 -39 -25 0, clip,draft=false,height=0.28\textwidth]{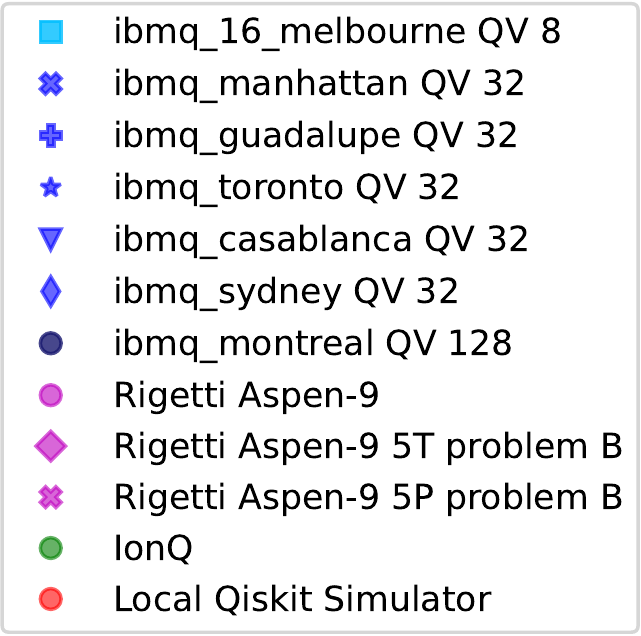}\hfill
    \includegraphics[trim = 0 0 0 0, clip,height=0.3\textwidth]{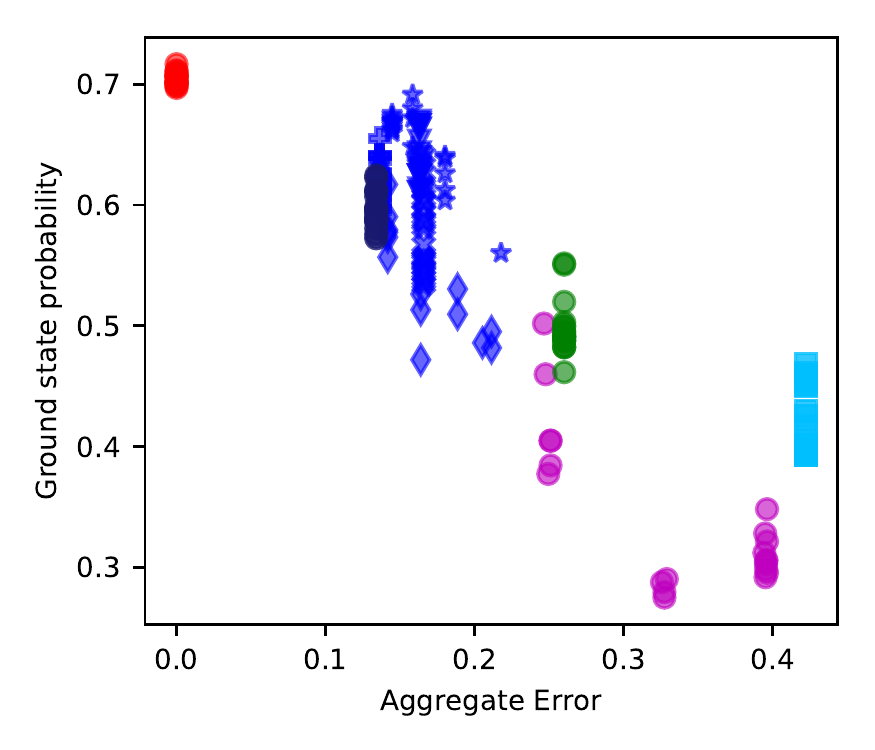}\hfill%
    \includegraphics[trim = 20 0 0 0, clip,height=0.3\textwidth]{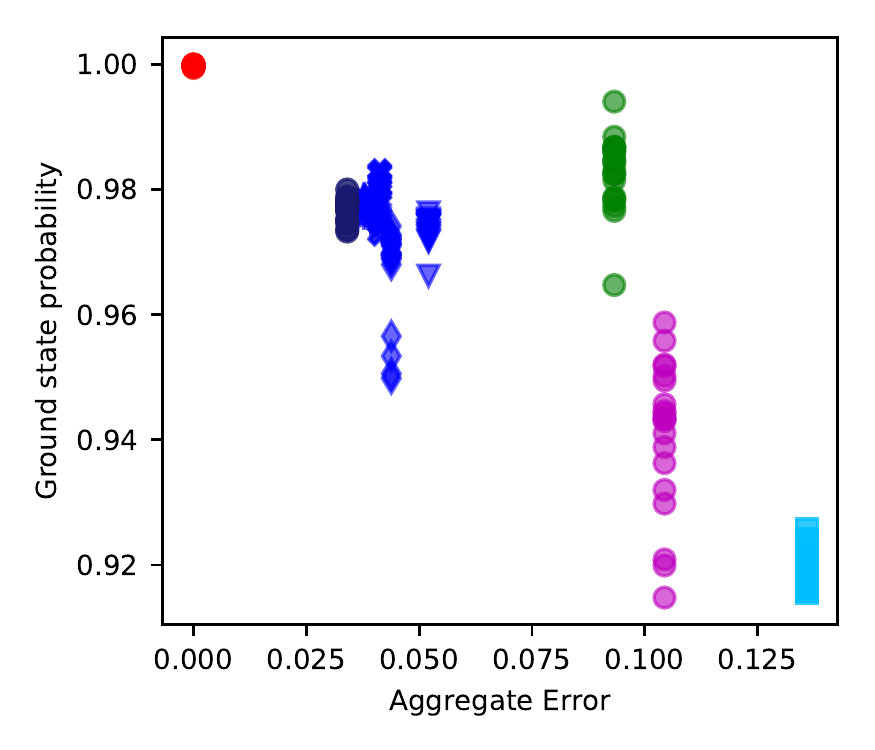}
    \caption{Aggregate error vs GSP for Problems (a) (top left), (b) (top center), (c) (top right), (d) (bottom center), (e) (bottom right).}
    \label{fig:results_aggregate_error_vs_GSP}
\end{figure*}

\begin{figure*}[t]
    \includegraphics[trim = 0 0 0 0, clip,draft=false, height=0.3\textwidth]{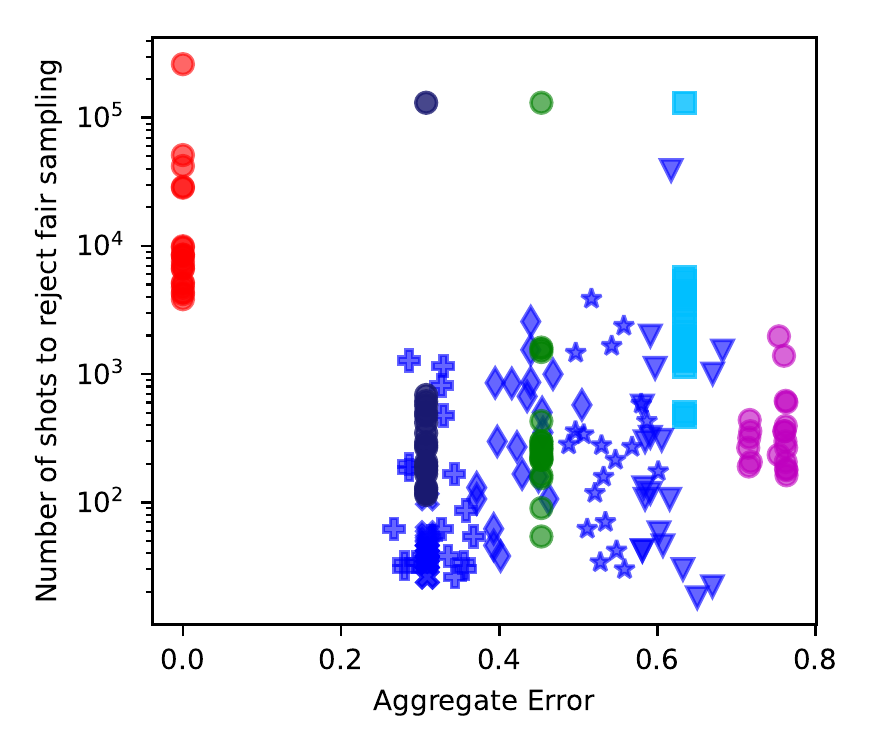}\hfill%
    \includegraphics[trim = 20 0 0 0, clip,draft=false, height=0.3\textwidth]{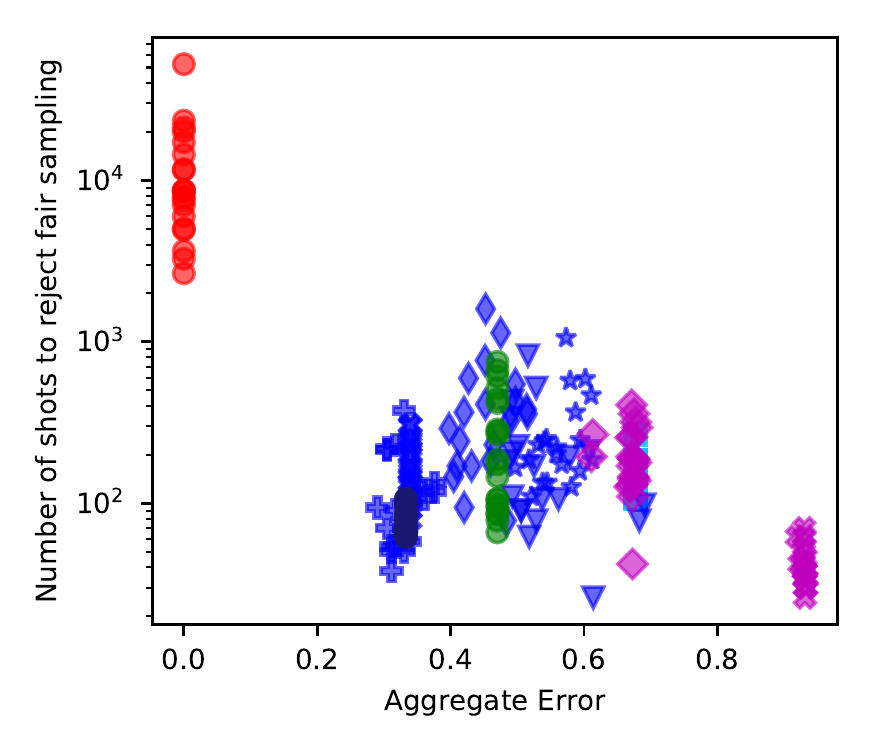}\hfill%
    \includegraphics[trim = 20 0 0 0, clip,draft=false,height=0.3\textwidth]{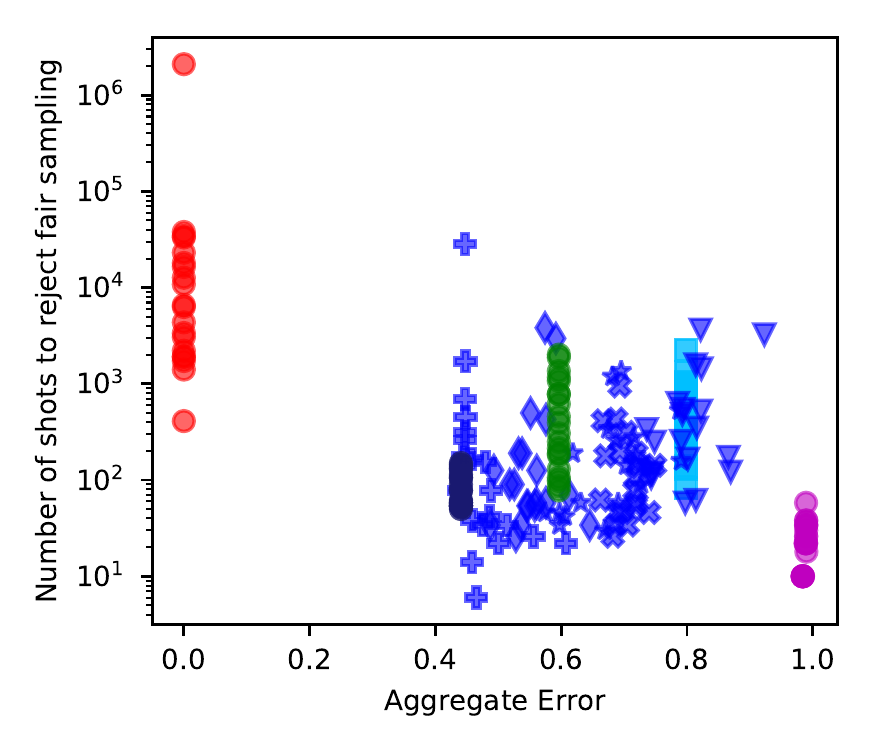}\\[-3ex]
    \hfill \includegraphics[trim = -47 -39 -24 0, clip,draft=false,height=0.28\textwidth]{figures/GSP_NSTR/legend_error.pdf}\hfill
    \includegraphics[trim = 0 0 0 0, clip,height=0.3\textwidth]{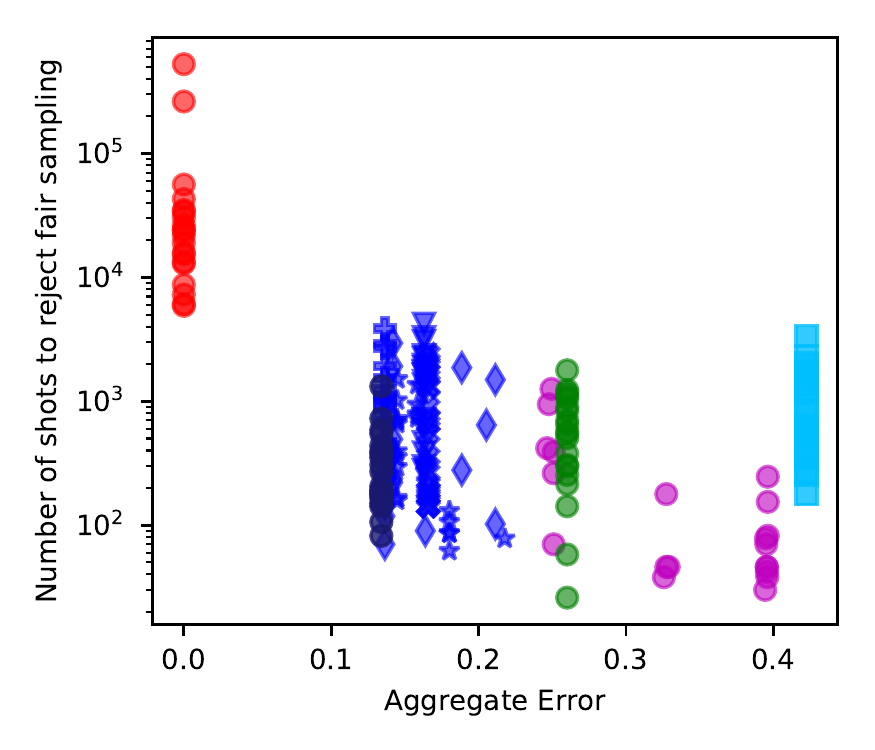}\hfill%
    \includegraphics[trim = 20 0 0 0, clip,height=0.3\textwidth]{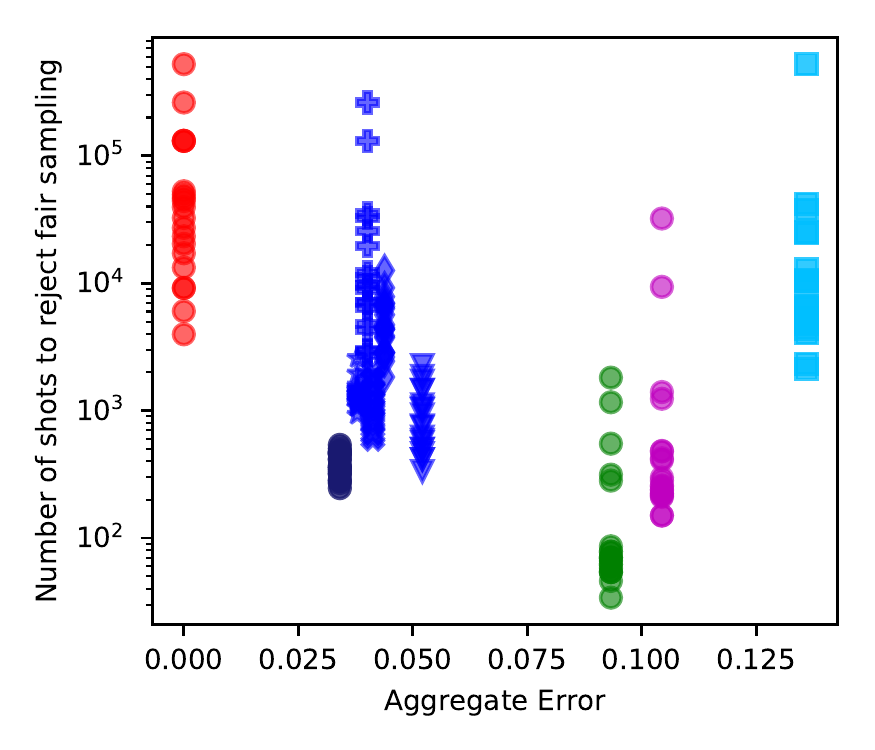}
    \caption{Aggregate error vs Fairness for Problems (a) (top left), (b) (top center), (c) (top right), (d) (bottom center), (e) (bottom right). }
    \label{fig:results_aggregate_error_vs_NSTR}
\end{figure*}

We briefly explain our experimental methodology and the experimental design. In general we send the vendor and backend-independent circuits from Section~\ref{sec:circuits} to the compiler for each of the vendors, calculate the aggregate error on each circuit, and then execute a large number of runs on the 12 backends in Table \ref{table:NISQ_Devices} with different circuit topologies described in Table \ref{table:topologies}, collecting data to calculate our metrics of ground state probability (GSP) and our fairness metric called ``number of shots to reject the fair sampling hypothesis''. We describe each of these individual steps briefly, particularly highlighting differences among the four vendors.

\begin{table}[ht]
	\centering
	\begin{tabular}{|c|c|c|c|c|}
		\hline
		& Backend & Type & QV & Qubits \\
		\hline
		IBM & melbourne & Gate model & 8 & 15\\
		\hline
		IBM & casablanca & Gate model & 32 & 7\\
		\hline
		IBM & guadalupe & Gate model & 32 & 16\\
		\hline
		IBM & toronto & Gate model & 32 & 27\\
		\hline
		IBM & sydney & Gate model & 32 & 27\\
		\hline
		IBM & manhattan & Gate model & 32 & 65\\
		\hline
		IBM & montreal & Gate model & 128 & 27\\
		\hline
		IonQ & ionq & Gate model & & 11\\
		\hline
		Rigetti & Aspen-9 & Gate model & & 31\\
		\hline
		DWave & LANL 2000Q & Annealer & & 2032\\
		\hline
		DWave & 2000Q-6 & Annealer & & 2048\\
		\hline
		DWave & Advantage & Annealer & & 5760\\
		\hline
	\end{tabular}
	\caption{Device properties for the NISQ backends we studied. Only IBM~Q uses the Quantum Volume metric (QV) to characterize their systems.}
	\label{table:NISQ_Devices}
\end{table}

\subsection{Backend Compilation}
\subsubsection*{IBM~Q}
\label{sec:methods_IBMQ}
We selected seven IBM~Q backends shown in Table \ref{table:NISQ_Devices}, which all allowed us to embed all five fair sampling problems onto the hardware connectivity, that is without having to resort to extensive swapping. 
We implemented the circuits of the five test problems in IBM's QISKIT (version 0.25.0). QISKIT has a method that allows you to compile the logical circuit onto the specified hardware using the native gateset. We run each circuit using three different compiler arguments: 
(i) simply leaving everything to default, except supplying the specific \textit{backend}; 
(ii)setting the \textit{layout method} as \textit{noise-adaptive}, setting the \textit{optimization level} to 3, default values otherwise; 
(iii) supplying an argument for \textit{initial-layout}, and set \textit{optimization level} to 3. The initial layout we supply is a random subgraph of the hardware connectivity graph which is \textbf{isomorphic} to the original circuit graph structure (e.g. 5T, 7H, etc). 
We used the circuit topologies as described in Table~\ref{table:topologies}.

\begin{figure*}[t]
    \includegraphics[trim = 0 0 0 0, clip,draft=false, height=0.3\textwidth]{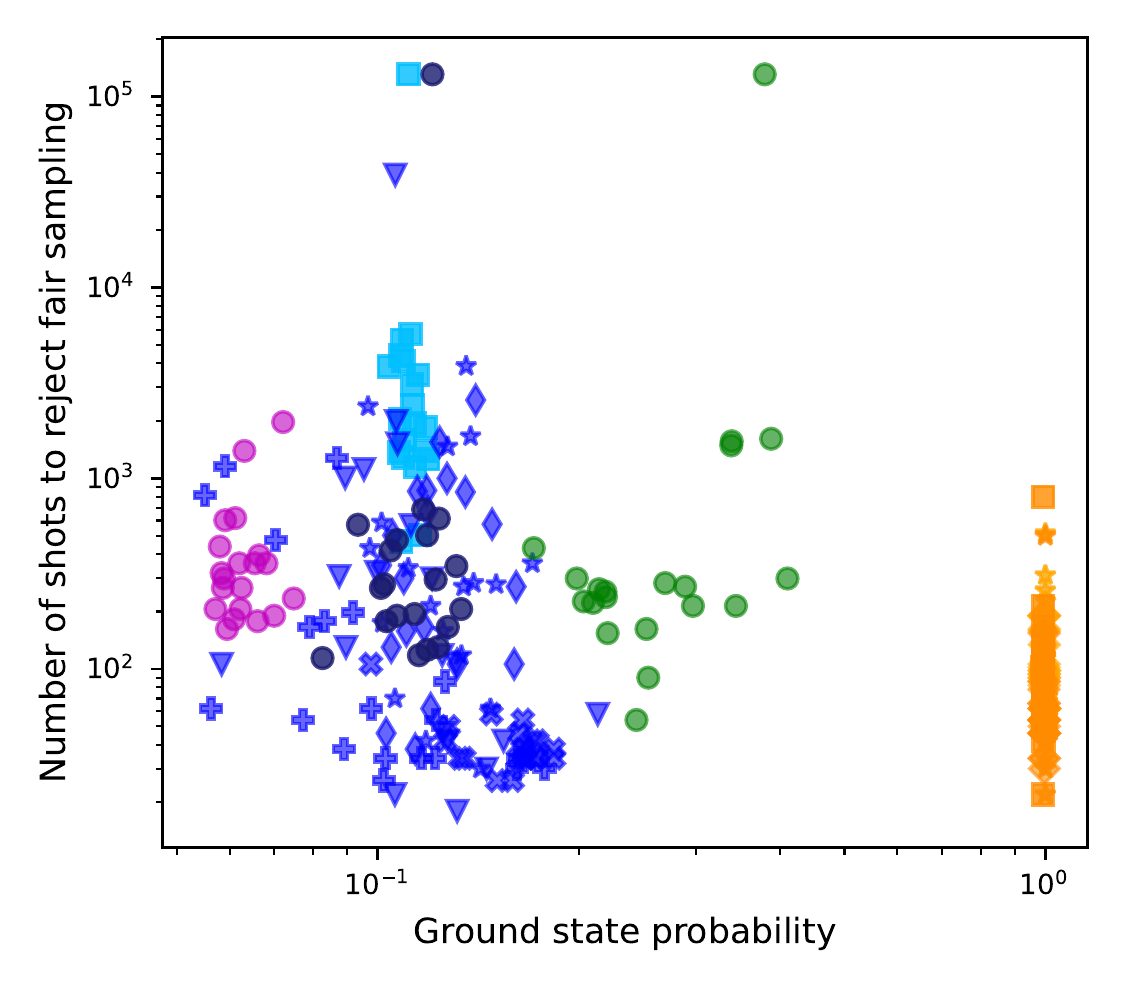}\hfill%
    \includegraphics[trim = 21 0 0 0, clip,draft=false, height=0.3\textwidth]{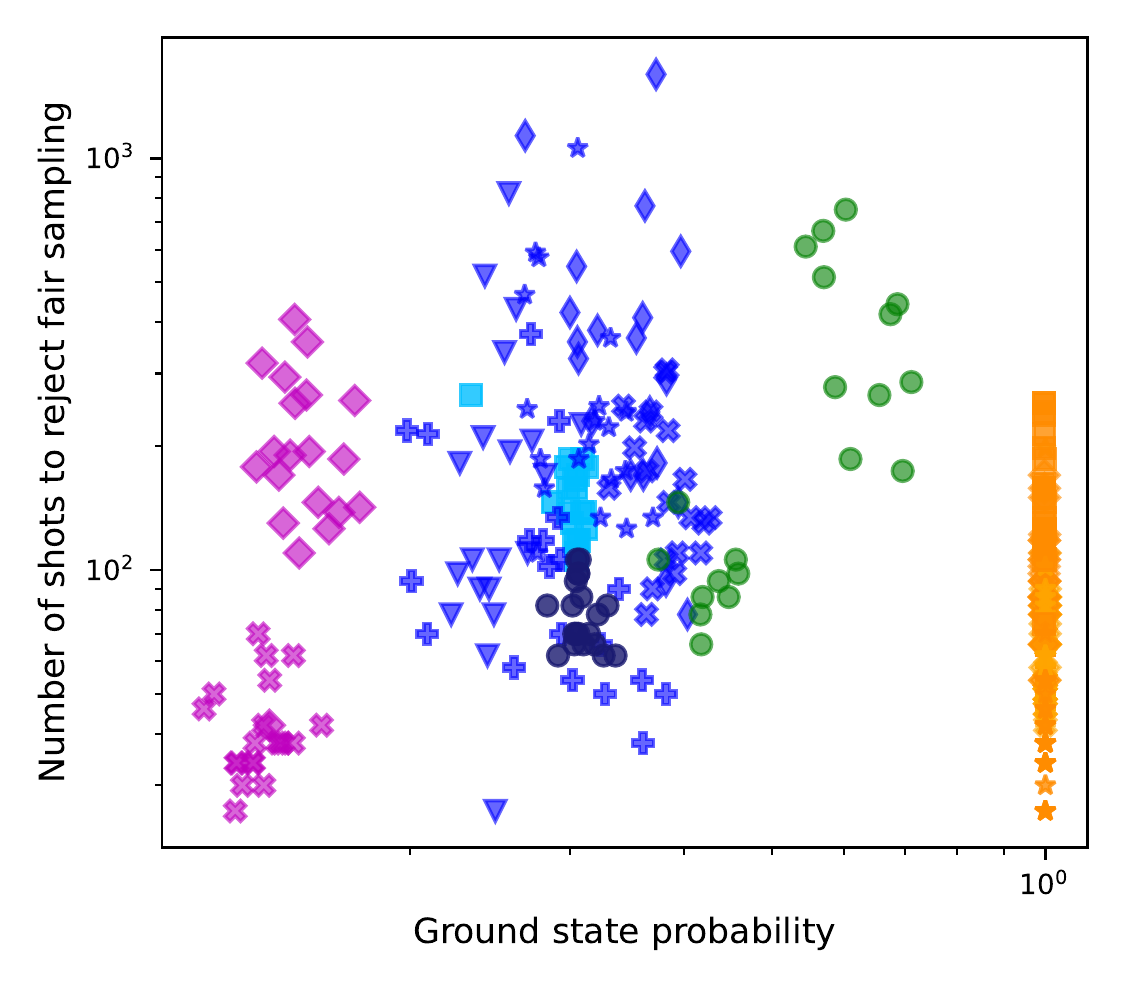}\hfill%
    \includegraphics[trim = 21 0 0 0, clip,draft=false,height=0.3\textwidth]{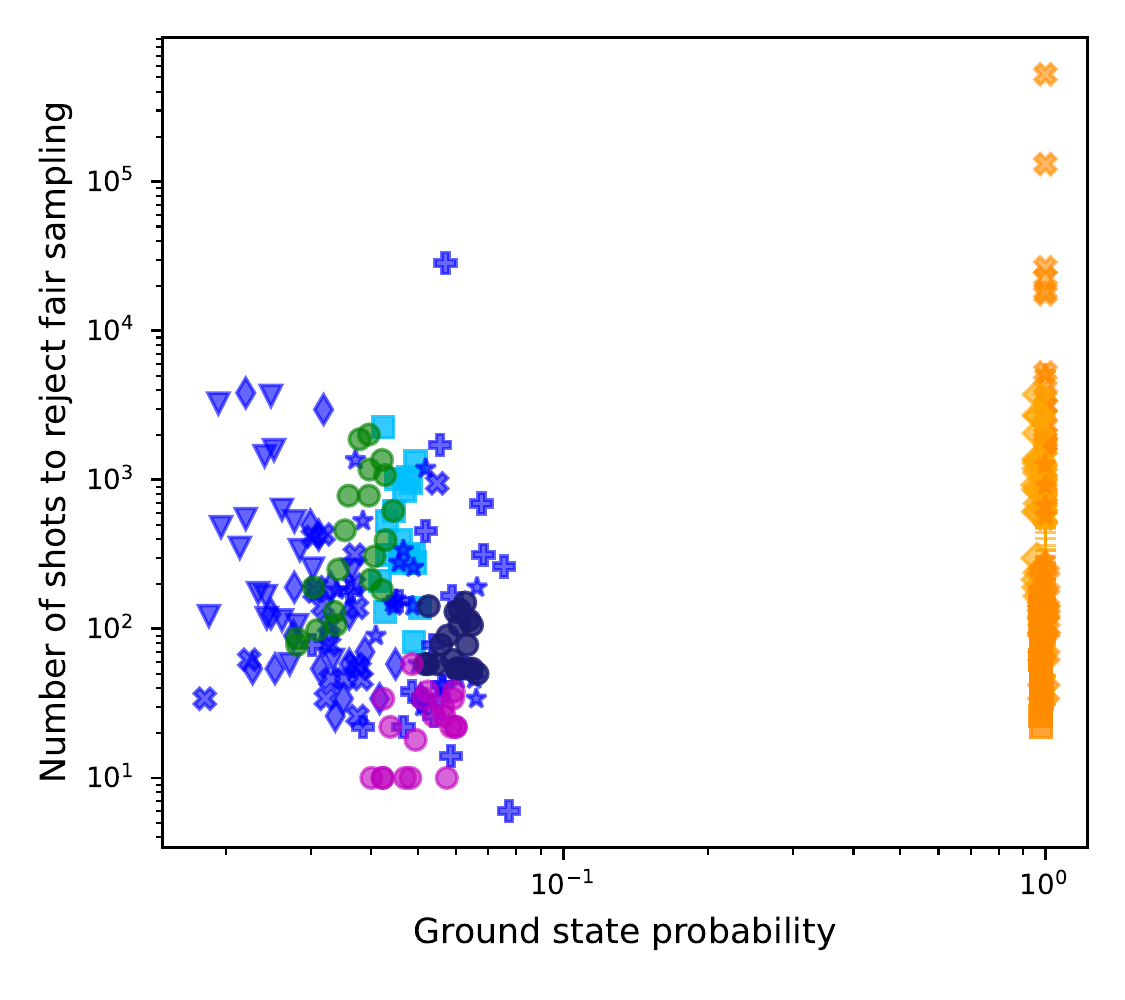}\\[-3ex]
    \hfill \includegraphics[trim = -52 -15 -5 0, clip,draft=false,height=0.28\textwidth]{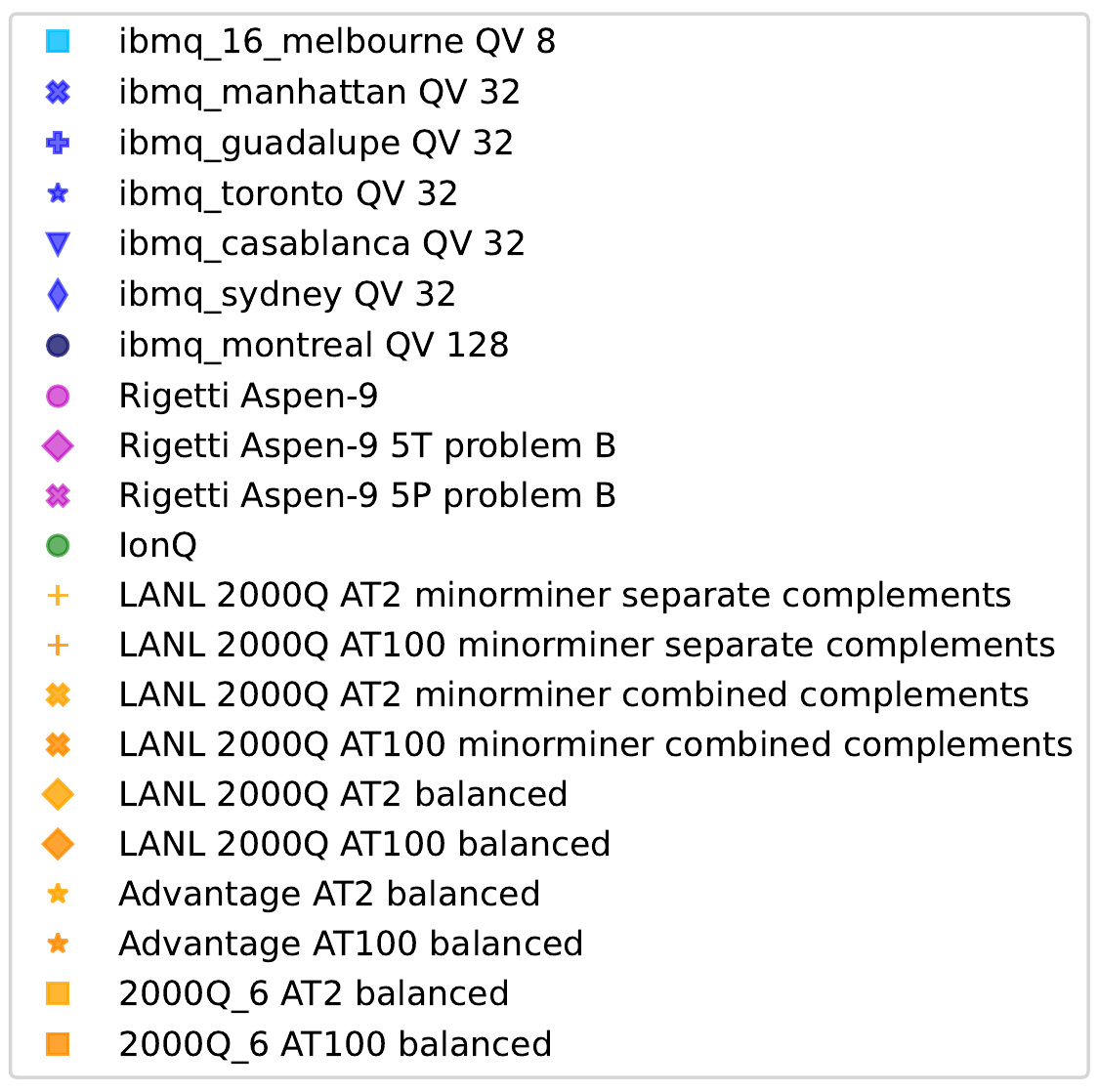}\hfill
    \includegraphics[trim = 0 0 0 0, clip,height=0.3\textwidth]{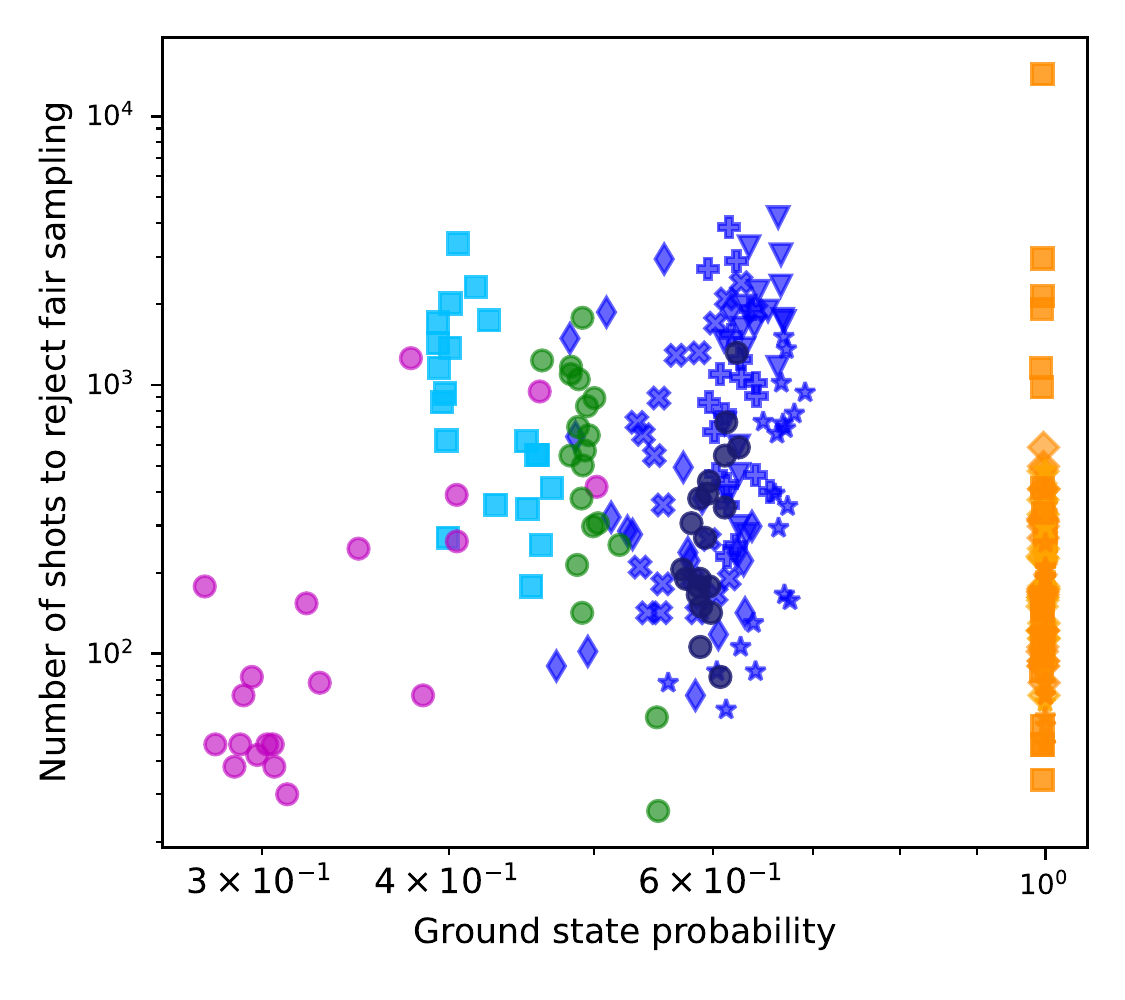}\hfill%
    \includegraphics[trim = 21 0 0 0, clip,height=0.3\textwidth]{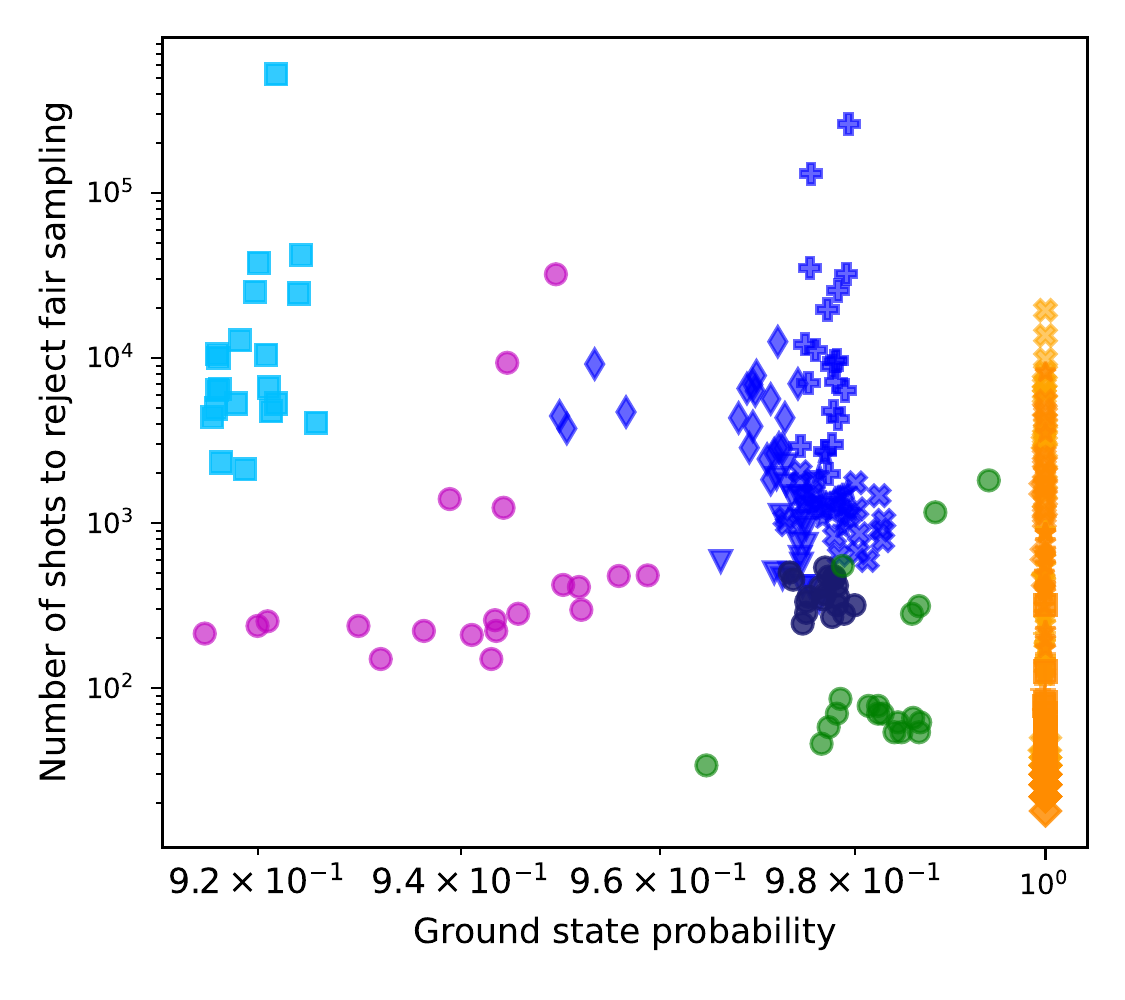}
    \caption{Log Ground state probability vs log Fairness for problems (a) (top left), (b) (top center), (c) (top right), (d) (bottom center), (e) (bottom right) across all 12 NISQ backends. }
    \label{fig:results_GSP_vs_NSTR}
\end{figure*}

\begin{figure*}
    \includegraphics[trim = 10 0 0 0, clip, draft=false, height=0.21\textwidth]{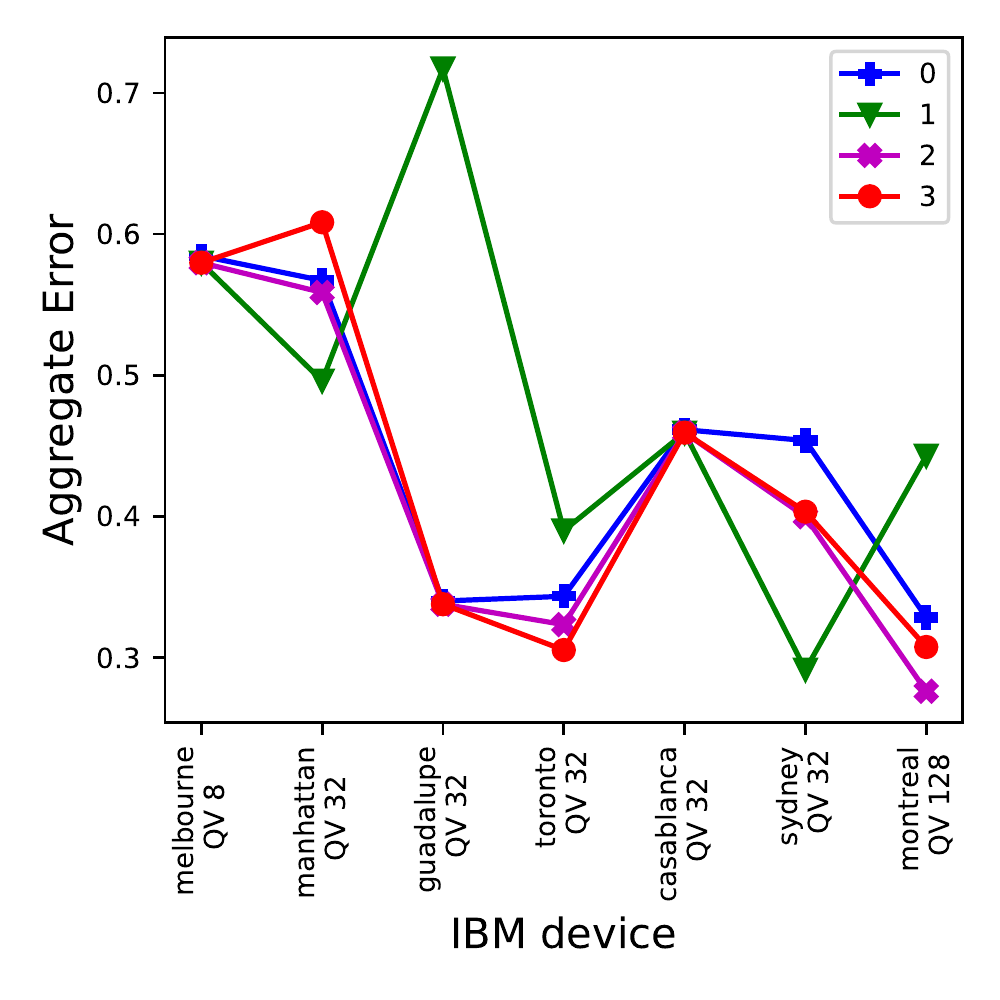}\hfill%
    \includegraphics[trim = 28 0 0 0, clip, draft=false, height=0.21\textwidth]{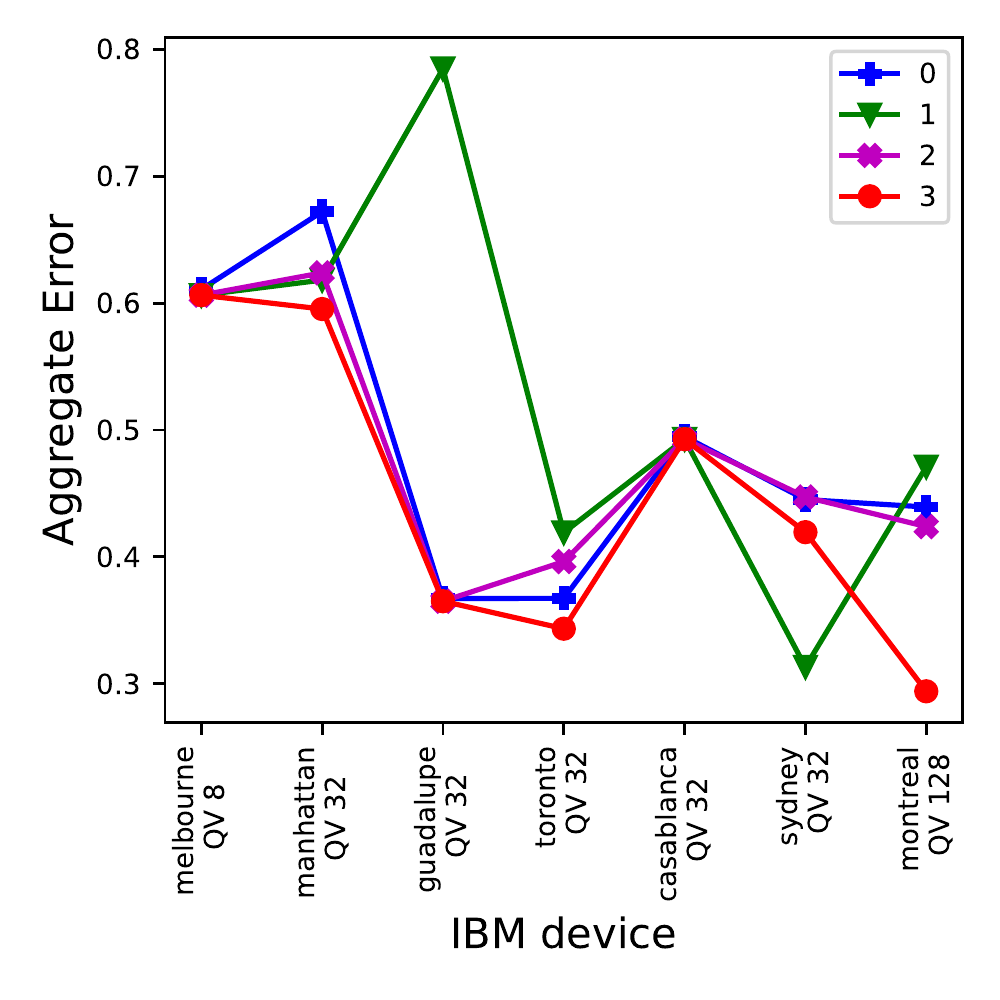}\hfill%
    \includegraphics[trim = 28 0 0 0, clip, draft=false, height=0.21\textwidth]{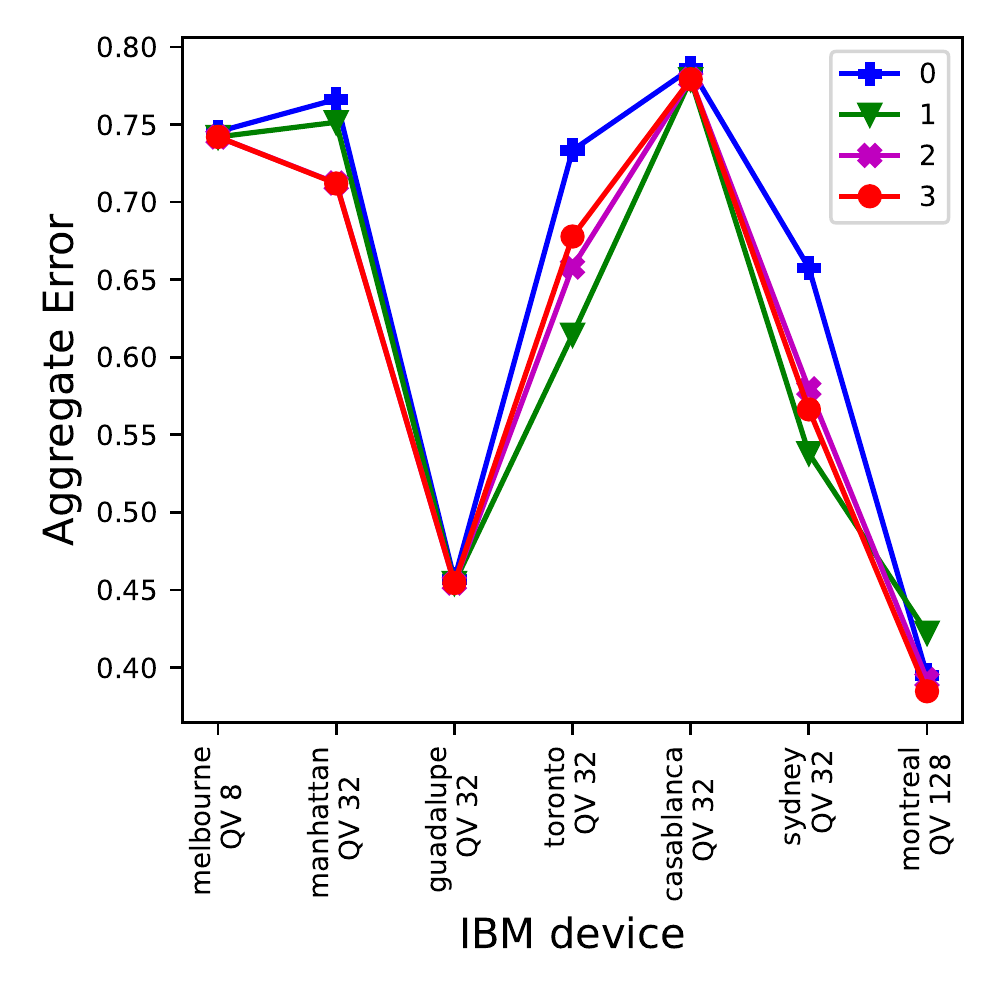}\hfill%
    \includegraphics[trim = 28 0 0 0, clip, draft=false, height=0.21\textwidth]{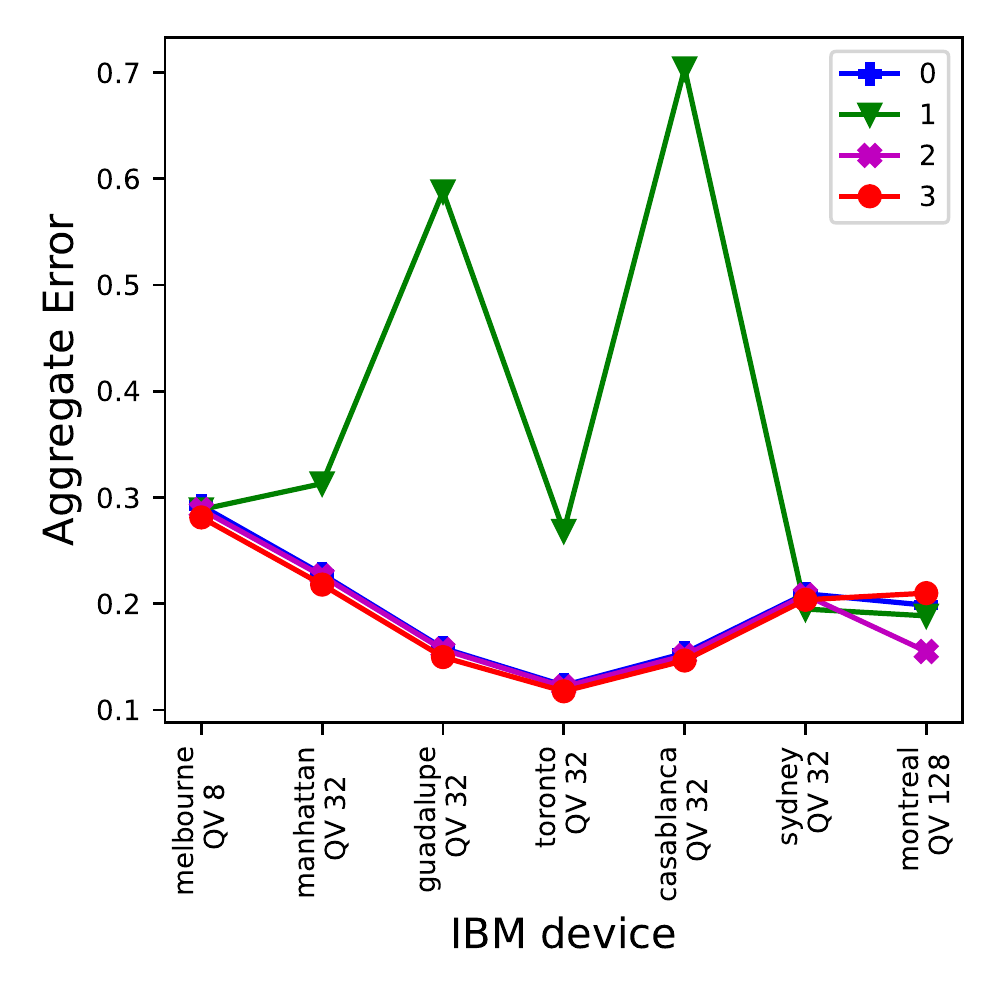}\hfill%
    \includegraphics[trim = 28 0 0 0, clip, draft=false, height=0.21\textwidth]{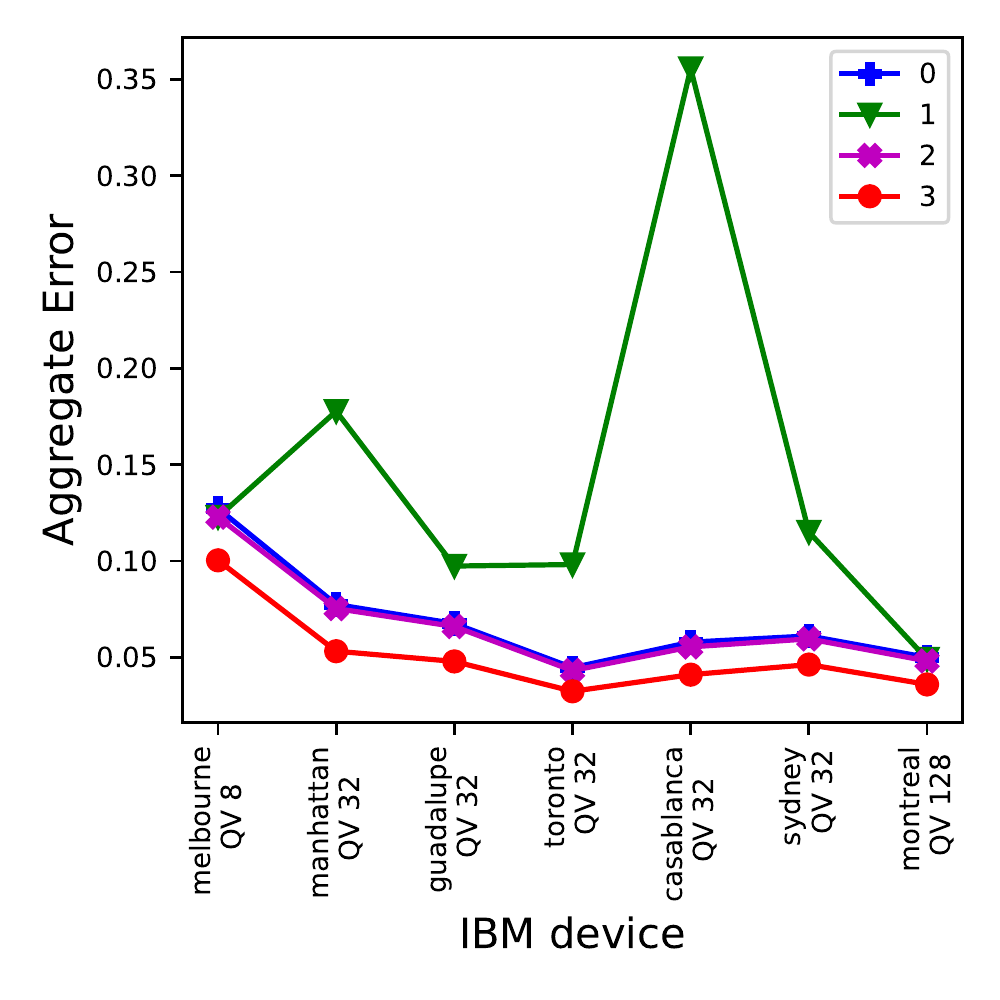}\\[-6.5ex]%
    \includegraphics[trim = 12 0 0 0, clip, draft=false, height=0.215\textwidth]{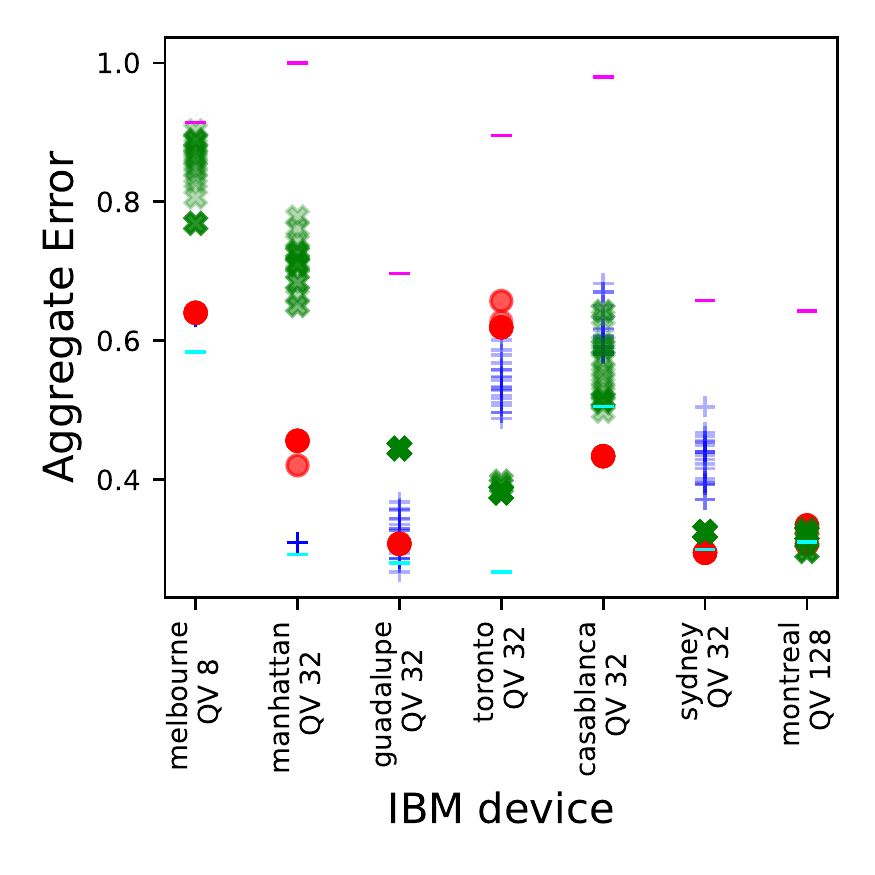}\hfill%
    \includegraphics[trim = 28 0 0 0, clip, draft=false, height=0.215\textwidth]{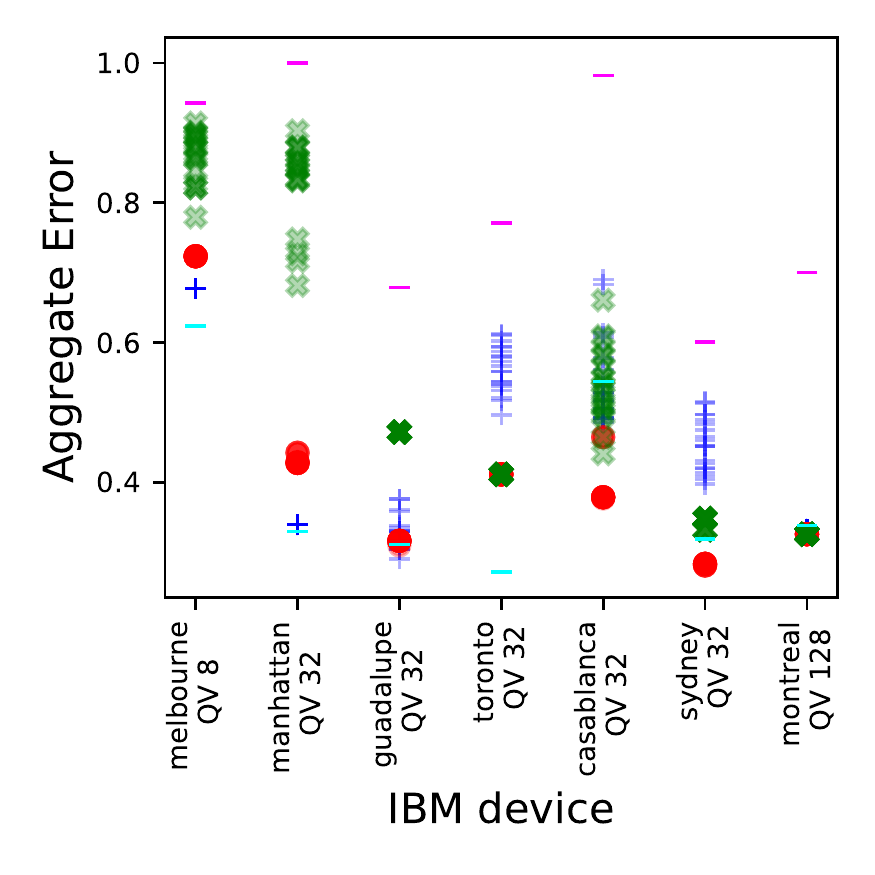}\hfill%
    \includegraphics[trim = 28 0 0 0, clip, draft=false, height=0.215\textwidth]{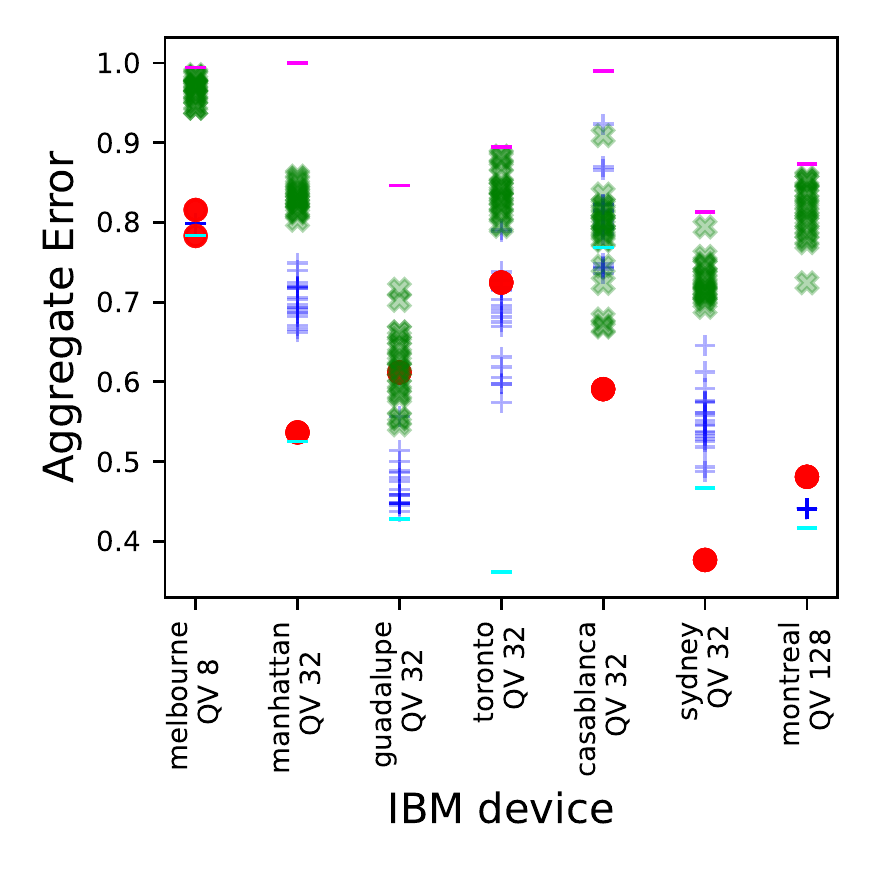}\hfill%
    \includegraphics[trim = 28 0 -5 0, clip, draft=false, height=0.215\textwidth]{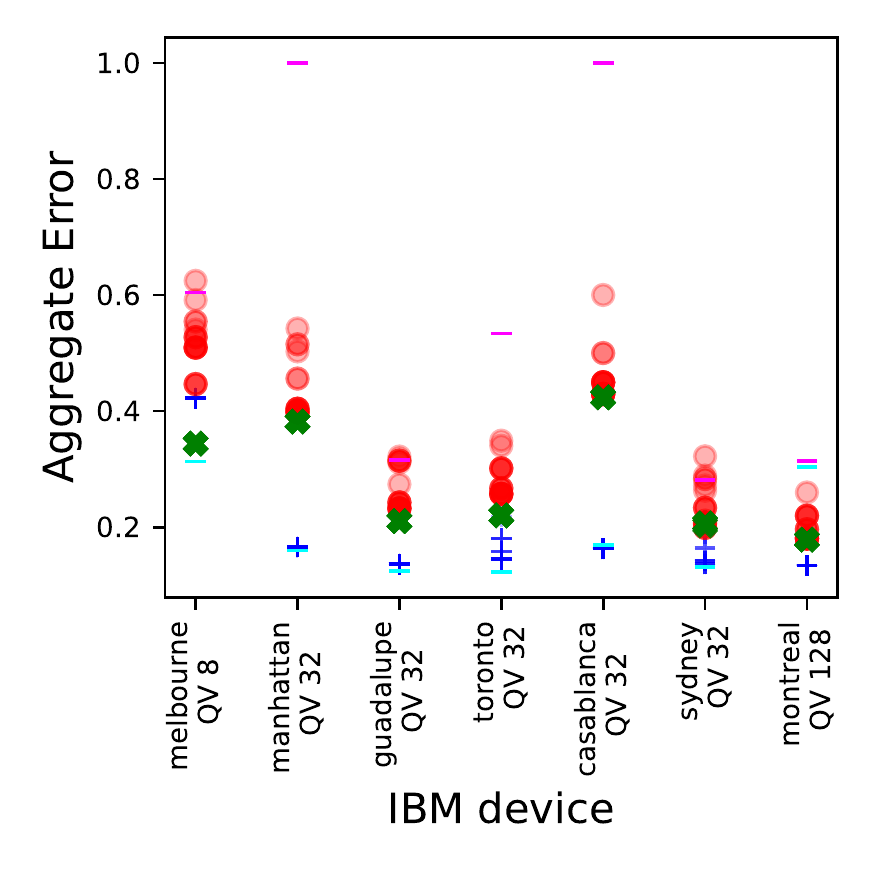}\hfill%
    \includegraphics[trim = 33 0 0 0, clip, draft=false, height=0.215\textwidth]{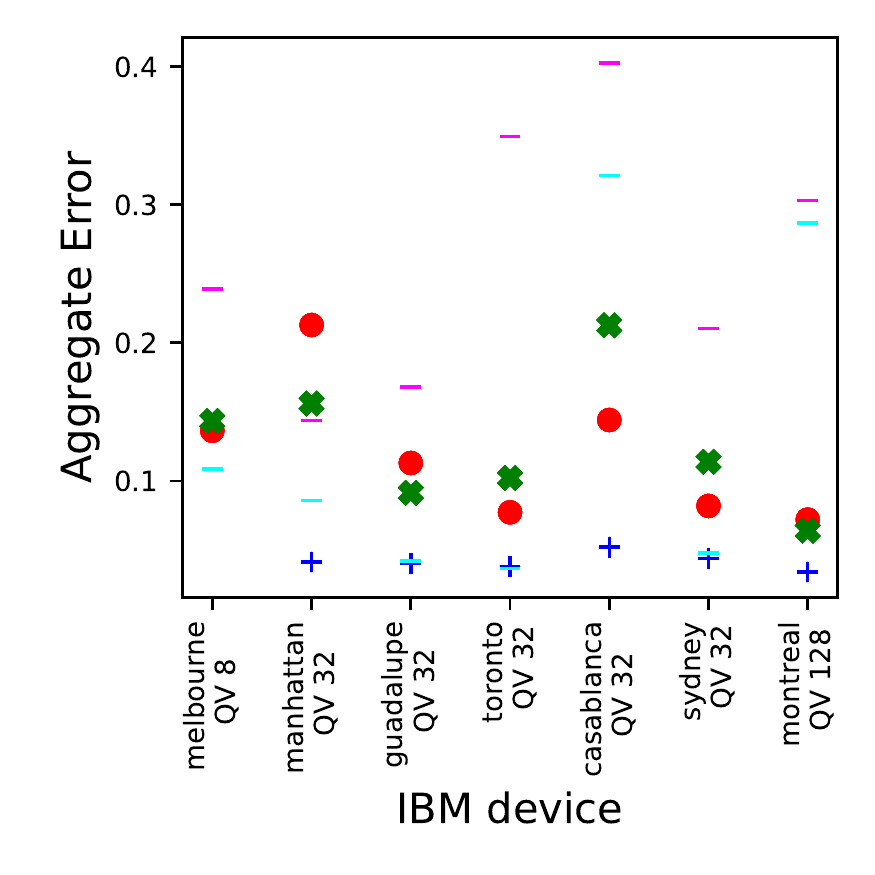}%
    \caption{IBM~Q device-specific aggregate errors for Problems (a), (b), (c), (d), (e) (from left to right) depending on transpilation:
        \textbf{(top)} We compare varying optimization levels 0, 1, 2, 3 for transpilation with noise adaptive layout for the respective devices. 
        \textbf{(bottom)} We plot 3 different transpilation methods: 
        Default (\textcolor{OliveGreen}{$\times$}), Initial layout (\textcolor{red}{$\circ$}) and Noise adaptive (\textcolor{blue}{$+$}), 
        as well as minimum (\textcolor{cyan}{$-$}) and maximum (\textcolor{magenta}{$-$}) aggregate error over \emph{all possible} IBM transpiler flag combinations.
        Default is calling transpile with no additional flags. 
        Initial layout 3 specifies a randomly chosen layout isomorphic to the original circuit (with level 3 optimization). 
        Noise adaptive 3 uses a noise adaptive layout (with level 3 optimization).
    }
    \label{fig:results_IBM_transpilation}
\end{figure*}

\subsubsection*{Rigetti}
\label{sec:methods_Rigetti}
We accessed the IonQ system through the AWS Braket cloud service. AWS Braket defines its own Python-based quantum circuit language. In order to successfully run on the Aspen-9 device, we found it necessary to hand-compile some aspects of the backend independent circuits as AWS Braket did not transform the circuits supplied into the native gateset used by Rigetti.
Rigetti has a compilation step from the provided circuit into their own Quil language. For each circuit we submit through AWS Braket, we get that compiled Quil code as part of the task metadata. The compilation step involves re-mapping the logical circuit onto a set of physical qubits. We found that this compilation step significantly changes the structure of the original circuit. 
The circuit topologies that were run on the Aspen-9 device are given in Table \ref{table:topologies}. For problem (b), both 5T and 5P were embeddable onto the Aspen-9 hardware, so both topologies were used. We used the 6A circuit topology because 7H is not embeddable onto the Aspen-9 connectivity. 

\subsubsection*{IonQ}
\label{sec:methods_IonQ}
In order to access the IonQ system, we used AWS Braket. IonQ accepts a different gate set than Rigetti, so we had to again hand-compile the circuits. IonQ's all-to-all topology made this relatively easy. 

\subsubsection*{DWave}
\label{sec:methods_DWave}
The circuits we run on D-Wave are in the form of Ising models. Recall that Table \ref{tab:models} column 3 \textbf{Ising Hamiltonian $H_C$} shows the exact Ising models we use for these experiments. 

When using quantum annealing, if the Ising model connectivity does not embed onto the hardware connectivity, then we must minor-embed the Ising model onto the hardware graph. We use a D-Wave's standard embedding method called \textit{minorminer} as well as a hand-tuned embeddings.

When we use minor-embedding, we also need to use a parameter called \textbf{chain strength} in order to ensure that physical qubits representing a logical variable (we refer to these physical qubits as a chain) take the same value, (meaning they do not disagree on the logical variable state). If a chain of physical qubits disagrees, then this is called a \textbf{broken chain}. In order to resolve a broken chain, in our experiments we use the \textit{majority vote} method. Additionally, we compute chain strength using the default method in D-Wave's Ocean SDK.

For the standard embedding experiments, we use a random minor-embedding found by D-Wave's \textit{minorminer}~\cite{MM-software,MM-paper, minorminer}, with potentially unequal chain lengths. Using the full Isings means that the results we get from D-Wave include not only the ground states shown in the 3rd column of Table \ref{tab:models}, but also their complement, thus there are two ways of quantifying ground state bias: (i) treat the complement of each ground state as a separate ground-state; (ii). treat the complement of each ground state as being the same as the original. Call the first option \textit{separate ground-state complements} and the second option \textit{combined ground-state complements}. We ran standard embedding experiments only on the LANL D-Wave 2000Q. 

For manual embedding experiments, we designed \textit{balanced embeddings} across all devices and problems. In order to achieve this, we first fixed the value of qubit 0 on all problems to be 1 (note that for the gate model problems, we also fix the state of qubit 0). In this process, the quadratic terms ($q_0, q_i$ : $n$) associated with qubit 0 became linear terms on the neighboring qubit ($q_i$: $n$). Fixing qubit 0 to be 1 allowed us to then have chain length 2 embeddings for both 2000Q backends, and chain length 1 (i.e. native embedding) embeddings for Advantage-system1.1. Fixing qubit 0 to be 1 also causes the ground state behavior to change from above. Specifically, now there are no complement ground states; only the original groundstates shown in column 4 of Table \ref{tab:models} (except missing qubit 0 because we fixed its state). We ran this experiment on all three D-Wave backends. 

We also varied the \textit{annealing time} parameter in all our experiments. Specifically, for the LANL 2000Q we varied annealing time between 1 microsecond and 300 microseconds in step size of 1 microsecond. The hardware does not allow for annealing times less than 1 microsecond. 

For the backends Advantage-system1.1 and 2000Q-6, we used AWS Braket to execute the tasks, and therefore getting large numbers of results was more limited. As such, we tried an annealing time of 2 microseconds, and an annealing time of 100 microseconds on both of these backends.

\subsection{Ground state probability metric}
\label{sec:methods_GSP}
Our first metric is simply \textit{ground state probability} (abbreviated as GSP), which is the fraction of shots which return a ground state solution for that circuit. 

\subsection{Fairness: Number of shots to reject fair sampling metric}
\label{sec:methods_NSTRFS}
We define our main metric, as the \textit{Number of shots to reject fair sampling} (for short simply called Fairness). We want to quantify how biased or fair the distribution of these found ground states is. Importantly, each of the five test problems has at least three ground state solutions, meaning that we can calculate a distribution of how frequently each ground state solution was found. 
The exact description of this method is outlined in \cite{golden2021qaoa}. The goal of this method is to provide a reasonable metric for fairness which translates p-values from the $\chi^2$ test into a larger (and more human-readable) metric. This metric specifically reports the number of shots (drawn from the ground state distribution) one would need in order to reject the fair sampling hypothesis with 95\% statistical significance. Thus, the smaller the metric is, the less fair the ground state distribution is, and the larger it is the more fair the ground state distribution is. 

The only adjustable parameter for this method, other than the input distribution, is the number of inner loops (called $n_i$). For all gate model results, we use $n_i = 100,000$ when applying the method, and for all D-Wave quantum annealing results we use $n_i = 1,000$ in order to reduce computation time.

\subsection{Aggregate Error metric}
\label{sec:methods_aggregate_error}
In order to quantify how much noise a particular circuit was subject to during it's execution, we used the \textit{aggregate error} metric introduced in \cite{golden2021qaoa}. To be specific, for a sequence of instructions (gates) $g_i, \dots, g_n$ that make up a circuit $C$, the \textit{aggregate error} is defined as 
$E_C = 1 - \prod_{i=1}^{n} (1-e_i)$ where $e_i$ is the associated error for the gate $g_i$. In this computation, we include every gate including the readout operation. 

In the case of IBM~Q, every task includes a full description of error rates for each native gate operation for each qubit and edge on the hardware. The calibration data for each backend is updated consistently (on the order of hours to days). Before submitting the task to each IBM~Q backend, we compiled the circuit into the native gateset for all IBM~Q devices using the QISKIT transpiler, see more details in Section \ref{sec:methods_IBMQ}. Therefore, we used these device compiled circuits in order to compute the \textit{aggregate error} for each circuit ran on the IBM~Q devices. 

In the case of Rigetti, which we accessed through AWS Braket, we queried the latest calibration data at the same time we were executing each task. This calibration data seems to be updated regularly (on the order of hours or day). The Rigetti calibration information is similar to the IBM~Q data, except that they provided single qubit gate fidelities as an entire class (instead of differentiating between different gate's) and the data was given as gate fidelities instead of error rates. For each task submitted, we also get the device compiled Quil code for that circuit (see Section~\ref{sec:methods_Rigetti}). We then used that Quil circuit (and the queried calibration data) in order to compute aggregate error.

IonQ provides less detailed information, but does give mean two qubit gate fidelity, a mean single qubit fidelity, and a mean readout fidelity, which we used to calculate aggregate error for the IonQ circuits.

For the D-Wave quantum annealers we test, there is not a reported error metric. Therefore, we only use the aggregate error metric for the gate model devices. 

\section{Results}
\label{sec:results}

\subsection{Fair sampling comparison across IBM~Q, Rigetti, and IonQ}
\label{sec:results_compare_NISQ_results}

Because error is the main reason for falling short of theoretical performance, we first study how the two metrics GSP and Fairness change as a function of \textit{aggregate error} for the different gate model backends. Figure \ref{fig:results_aggregate_error_vs_GSP} shows how GSP changes for the five different test problems on each of the nine gate model backends. Similarly, Figure \ref{fig:results_aggregate_error_vs_NSTR} shows how Fairness changes across the nine gate model backends. We include points obtained from a classical Local QISKIT Simulator, which shows roughly what to expect expect from a zero-error quantum device. Across all figures in this subsection, we plot the metric values for each of the 20 individual calls (each call having 8192 shots) for each backend. Thus each plot in Figures \ref{fig:results_aggregate_error_vs_GSP}, \ref{fig:results_aggregate_error_vs_NSTR} and \ref{fig:results_GSP_vs_NSTR} shows at least 200 individual points plotted. For the IBM Q devices, the results plotted in Figures \ref{fig:results_aggregate_error_vs_GSP}, \ref{fig:results_aggregate_error_vs_NSTR} and \ref{fig:results_GSP_vs_NSTR} all used the noise adaptive transpiler options outlined in Section~\ref{sec:methods_IBMQ}. 

In theory, increasing aggregate error should correlate with decreasing values for both the GSP and Fairness. While both Figures \ref{fig:results_aggregate_error_vs_GSP} and \ref{fig:results_aggregate_error_vs_NSTR} show this general trend, there are fairly notable exceptions. The correct interpretation of these plots looks at the 20 points of each backend as a cloud of points; while we did not discard outlier points, we should take care to not interpret results solely based on outliers, but rather on the position of the cloud. 
With respect to GSP in Figure \ref{fig:results_aggregate_error_vs_NSTR}, IonQ reaches the best values in Problems (a), (b), and (e), despite having higher aggregate error than many of the IBM Q systems. Among the IBM Q system, there is substantially smaller aggregate error as we go from the older Quantum Volume (QV) 8 systems to the newer higher QV systems (light blue to dark blue). However, this reduced error does not consistently lead to improved GSP values, in fact several of the QV32 systems outperform the QV128 in most examples. IBM Q does however perform best for Problems (c) and (d), coming very close to the theoretical optimum (QISKIT simulator) in Problems (d) and (e). Rigetti's performance falls behind, largely due to high aggregate error.
With respect to fairness (Figure \ref{fig:results_aggregate_error_vs_NSTR}), we note the large variability in the Fairness metric that even extends to the simulator data. The high-error low QV IBM Q Melbourne system performs surprisingly well, most likely due to its unique topology, beating out its higher QV siblings, Rigetti and IonQ in Problems (a) and (e) and matching their performance in the other problems. The Rigetti system manages a similar performance coming close to IBM and IonQ in most problems despite high error. The QV128 IBM system again does not really stand out despite its generally low error rates. IonQ's fairness solidly matches IBM Q's in four problems, but falls well short in Problem~(e).

\begin{figure*}[t]
    \includegraphics[trim =  0 0 10 0, clip, draft=false, height=0.175\textwidth]{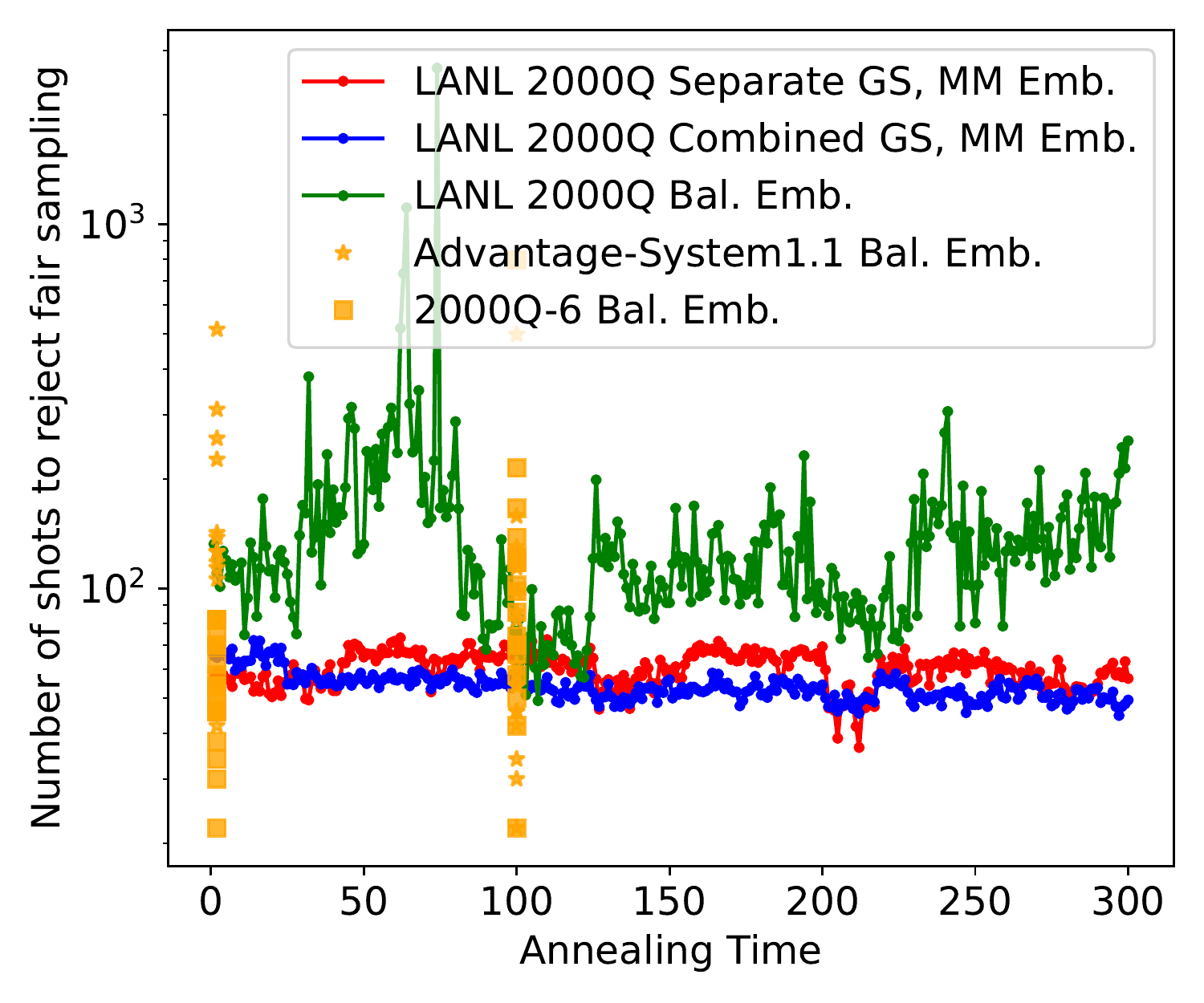}\hfill%
    \includegraphics[trim = 25 0 10 0, clip, draft=false, height=0.175\textwidth]{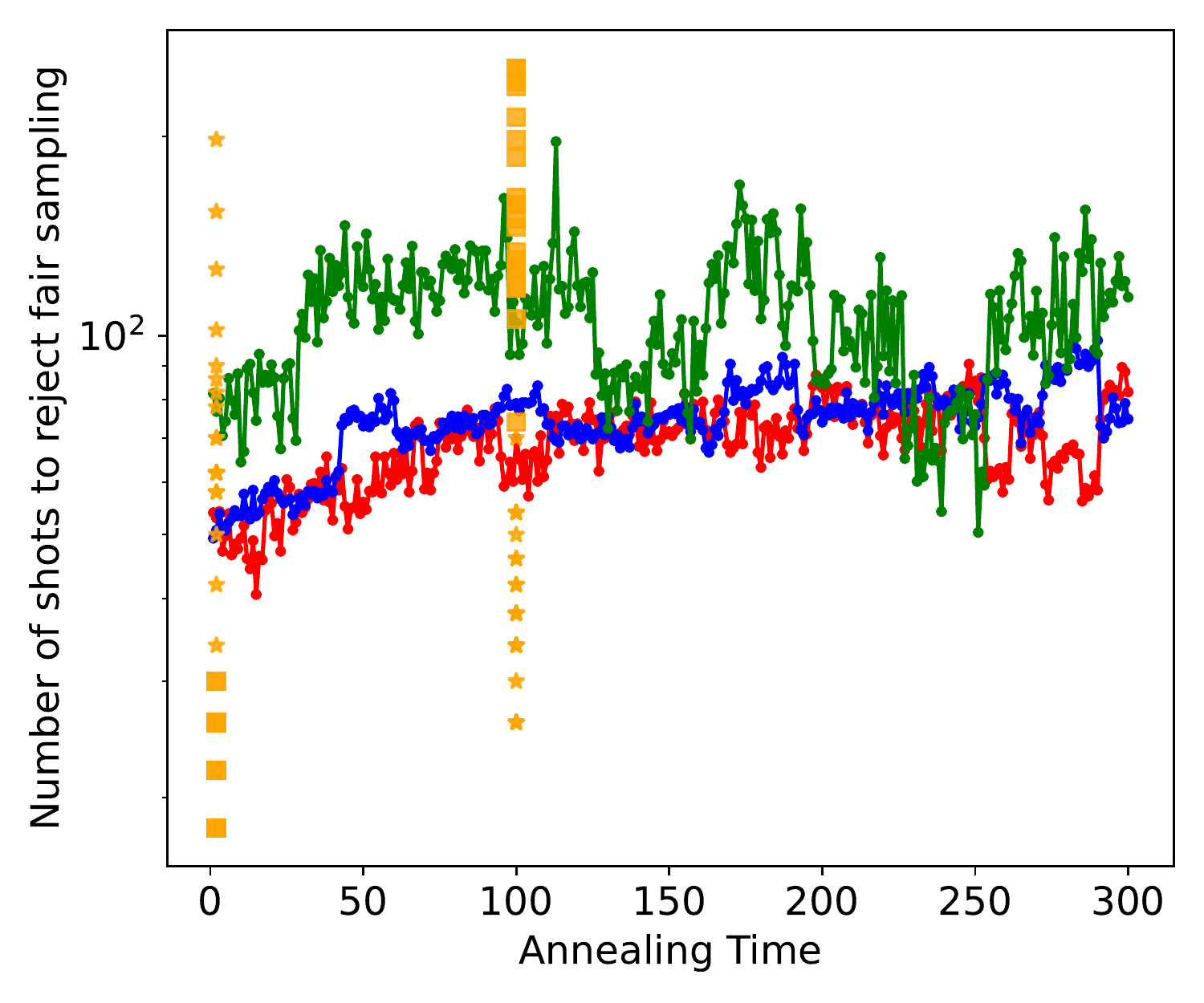}\hfill%
    \includegraphics[trim = 25 0 10 0, clip, draft=false, height=0.175\textwidth]{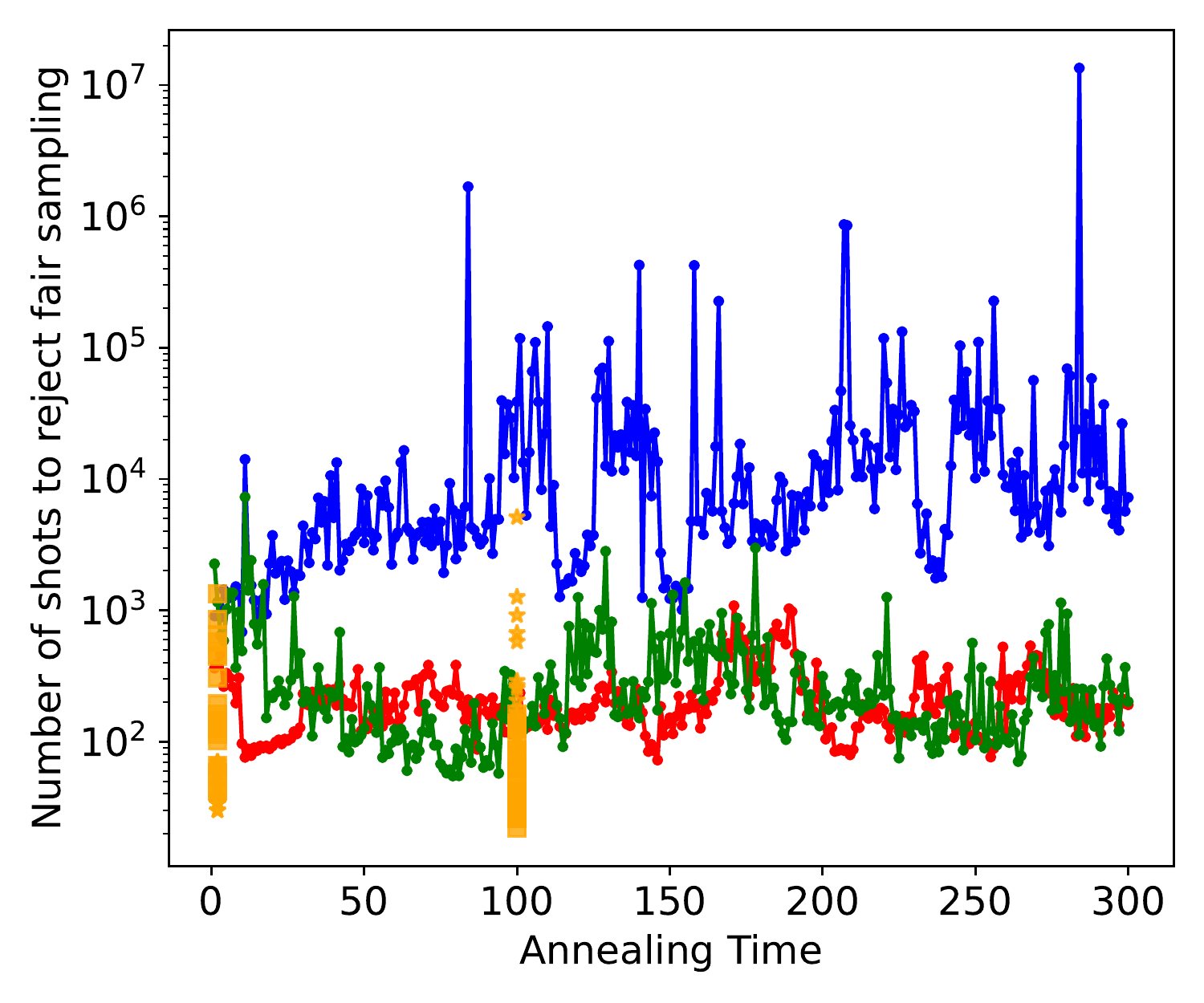}\hfill%
    \includegraphics[trim = 25 0 10 0, clip, draft=false, height=0.175\textwidth]{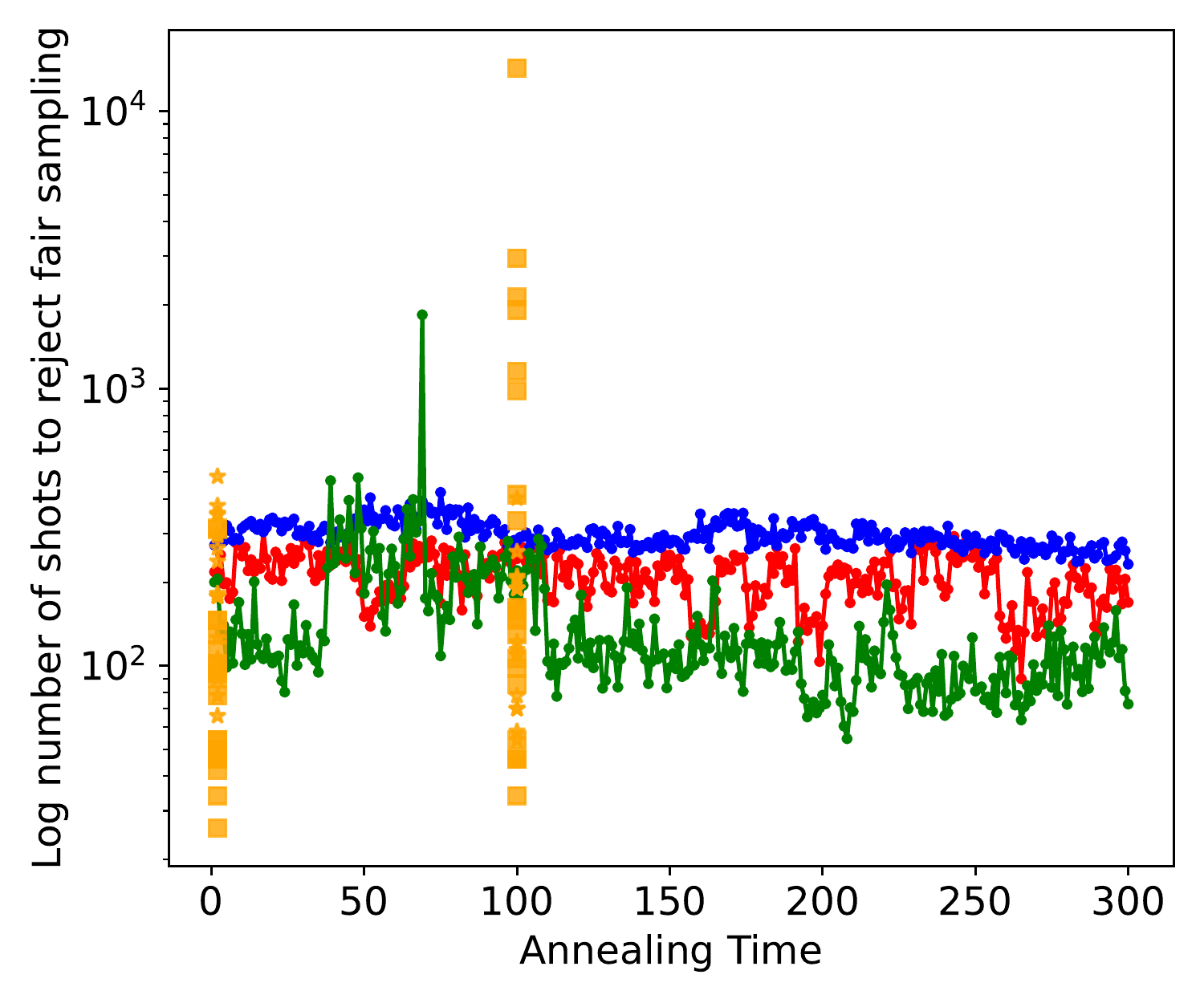}\hfill%
    \includegraphics[trim = 25 0 10 0, clip, draft=false, height=0.175\textwidth]{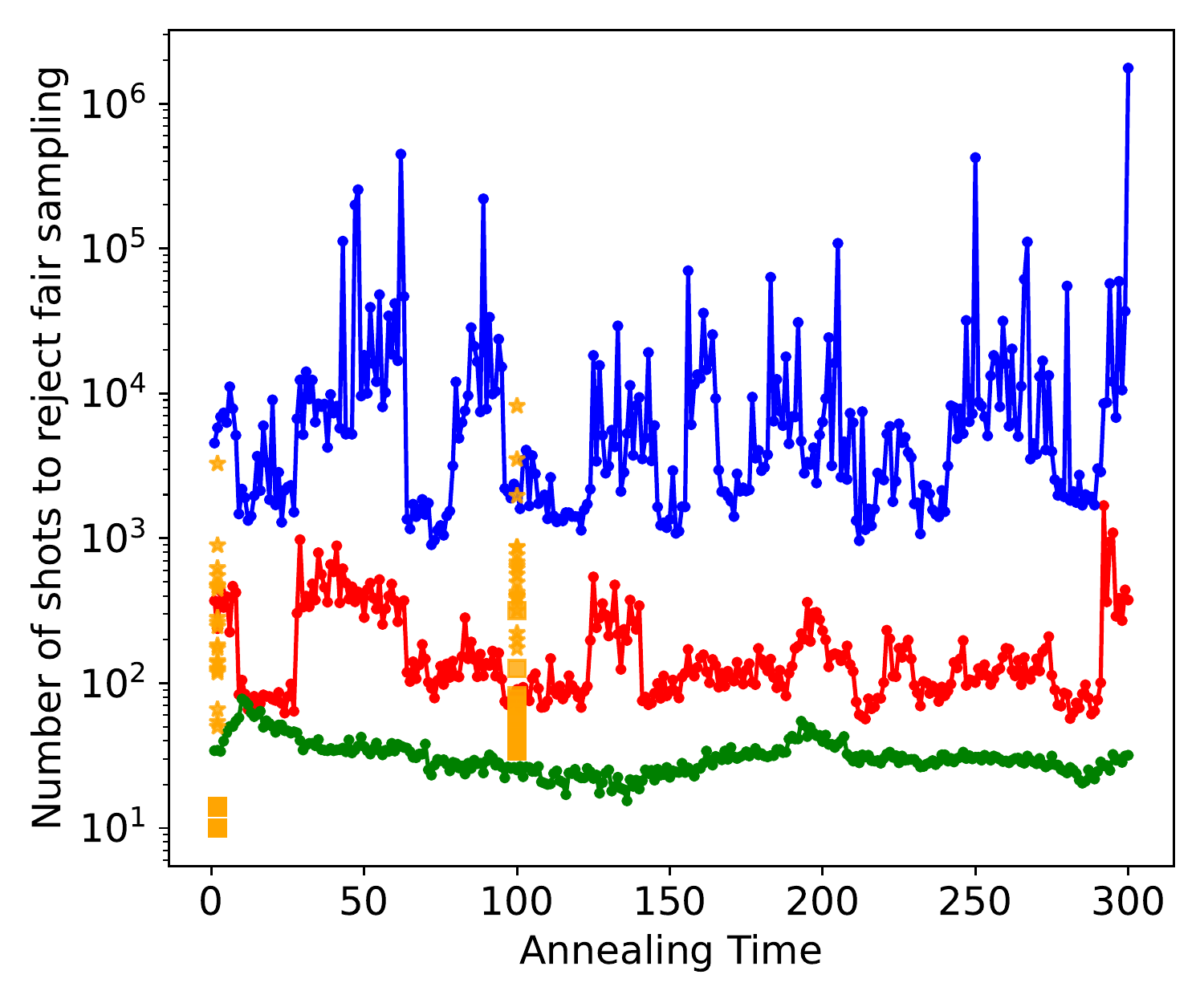}\\[-2.5ex]
    \includegraphics[trim =  0 0 10 0, clip, draft=false, height=0.175\textwidth]{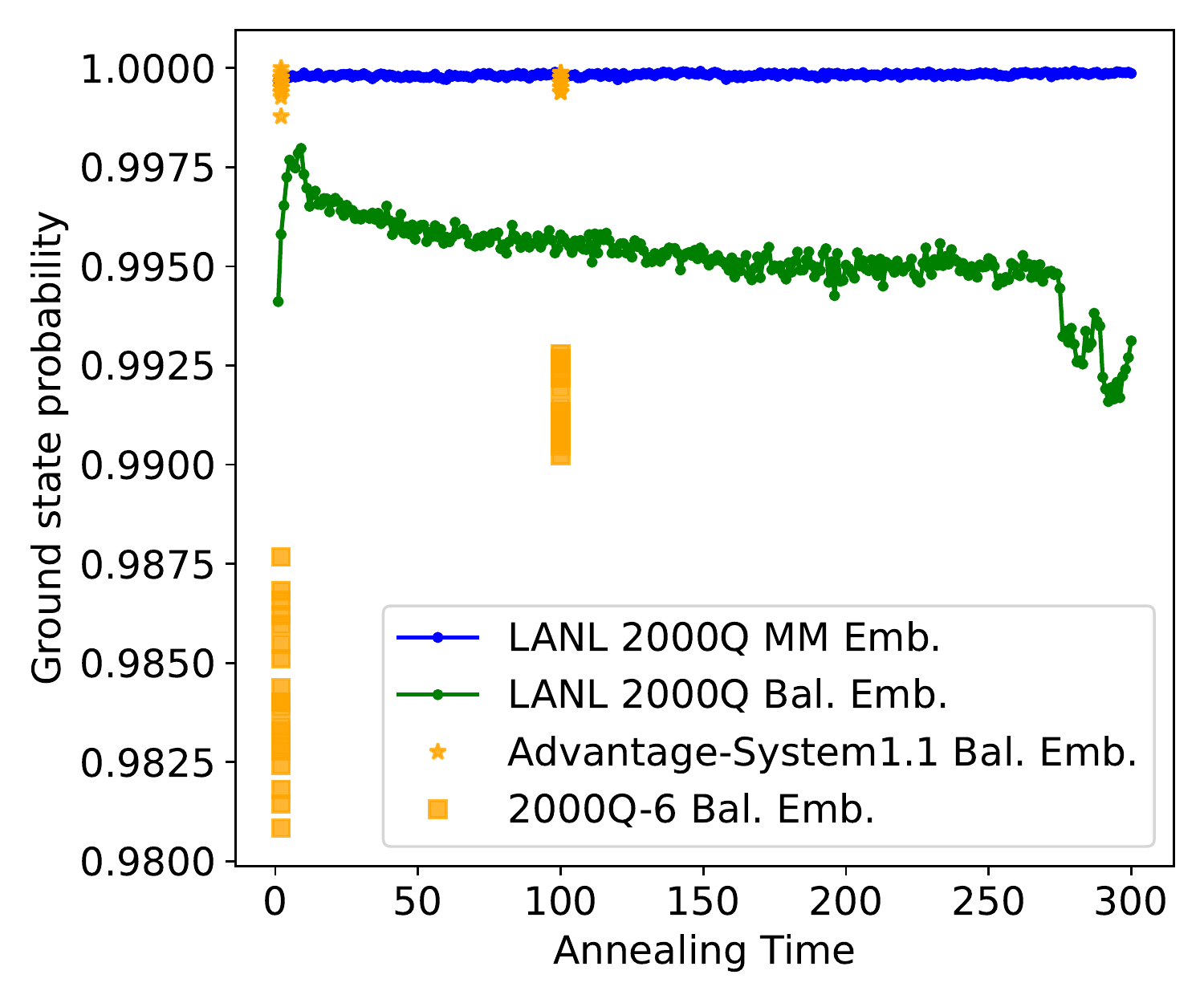}\hfill%
    \includegraphics[trim = 25 0 10 0, clip, draft=false, height=0.175\textwidth]{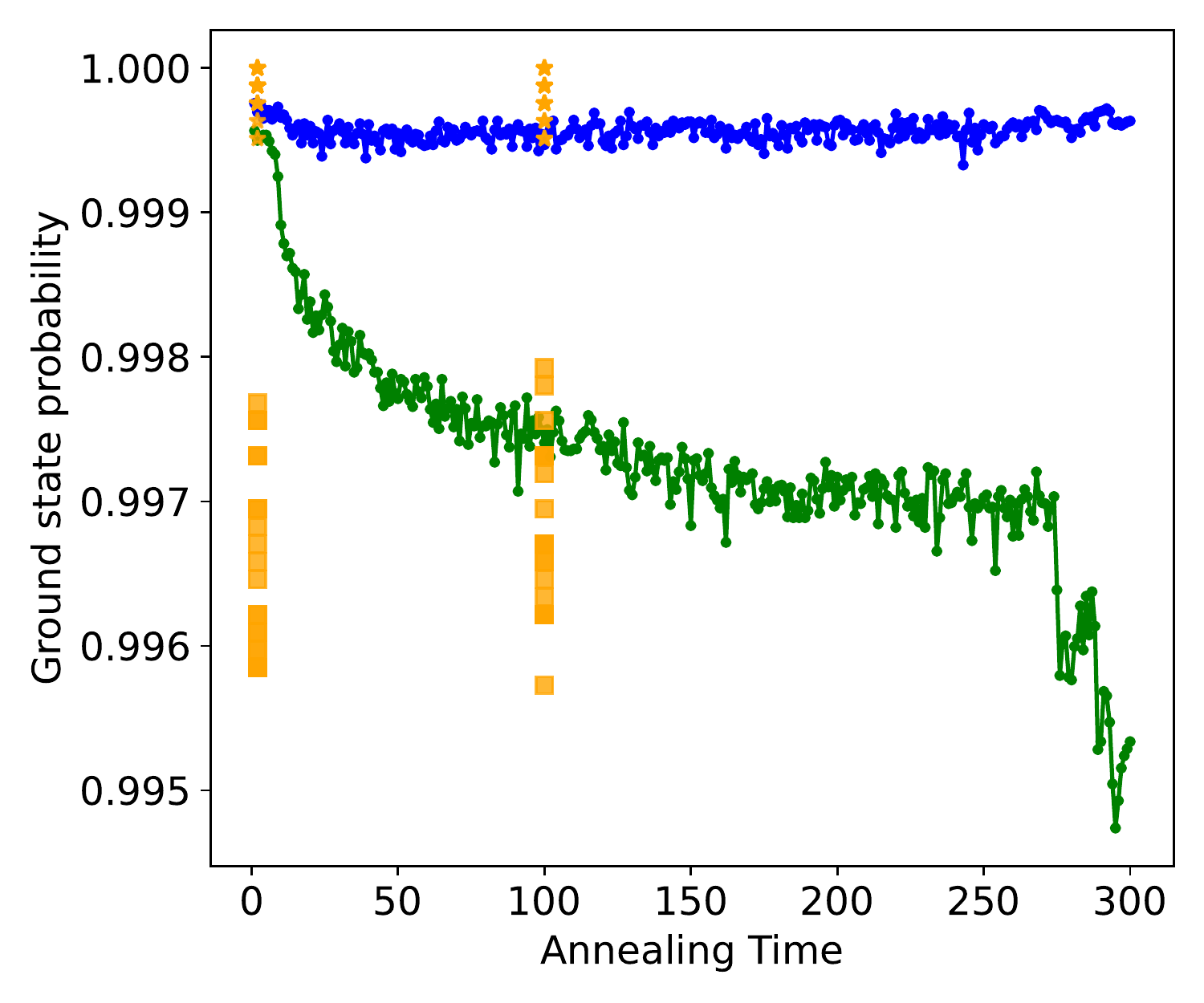}\hfill%
    \includegraphics[trim = 25 0 10 0, clip, draft=false, height=0.175\textwidth]{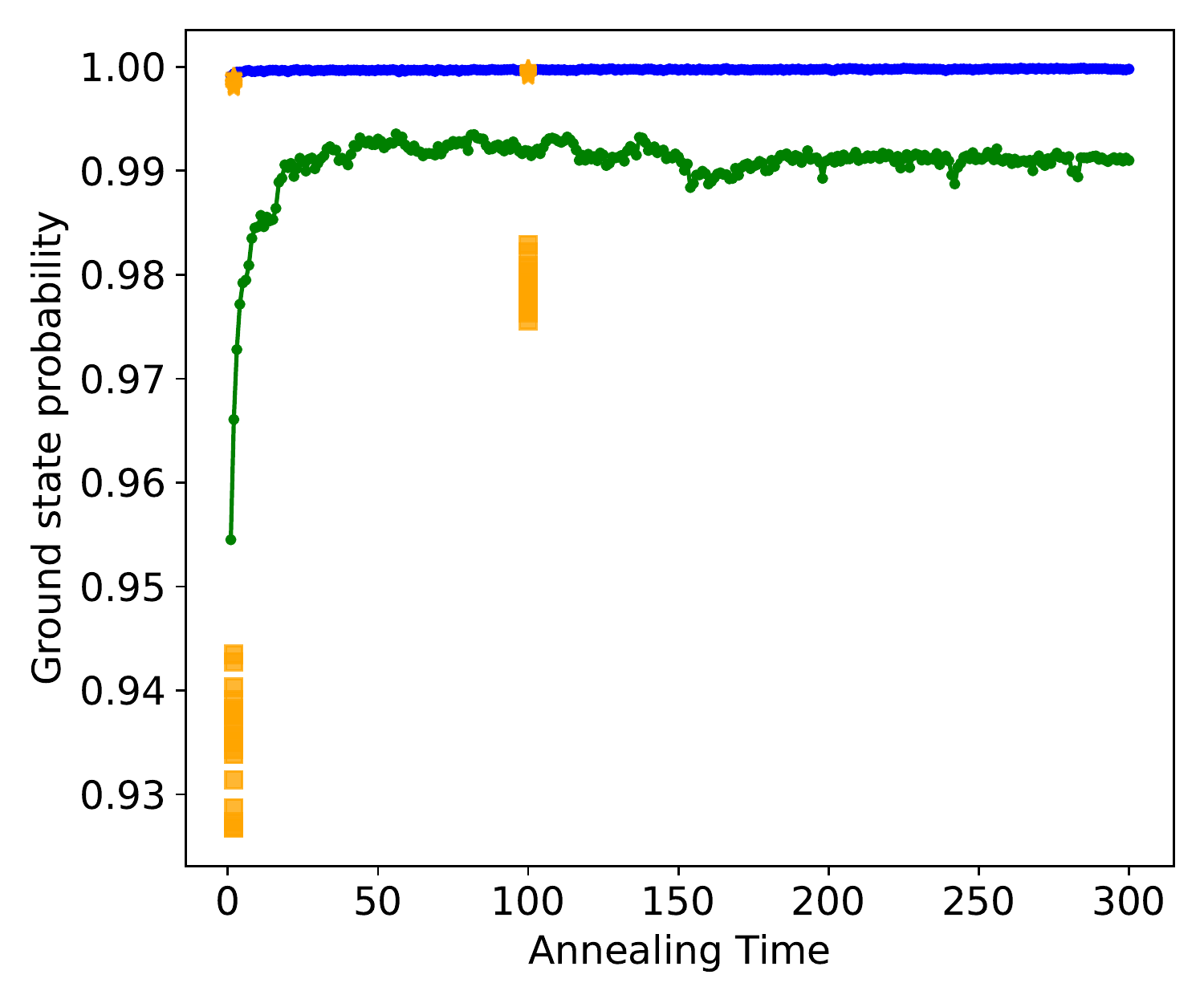}\hfill%
    \includegraphics[trim = 25 0 10 0, clip, draft=false, height=0.175\textwidth]{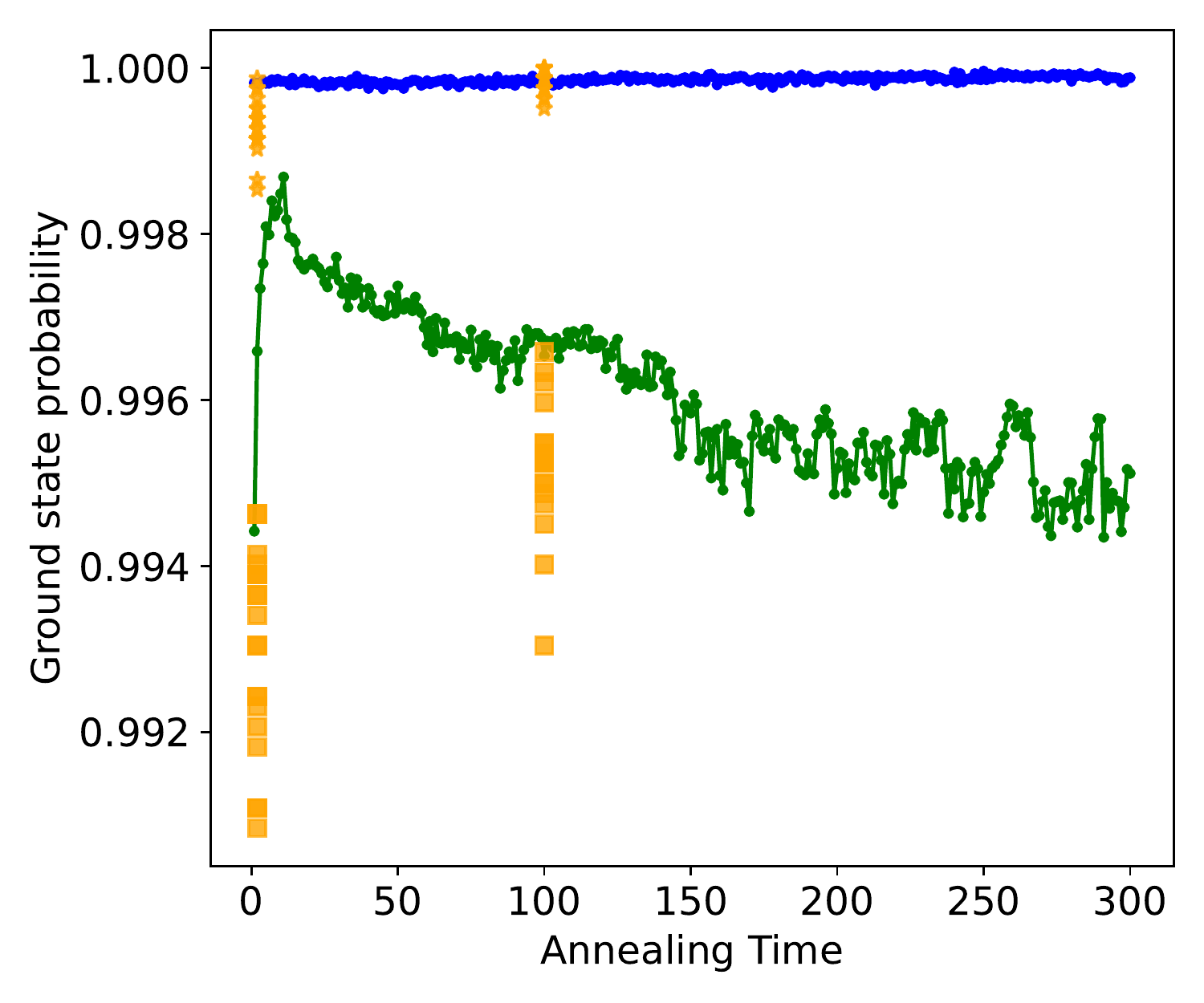}\hfill%
    \includegraphics[trim = 25 0 10 0, clip, draft=false, height=0.175\textwidth]{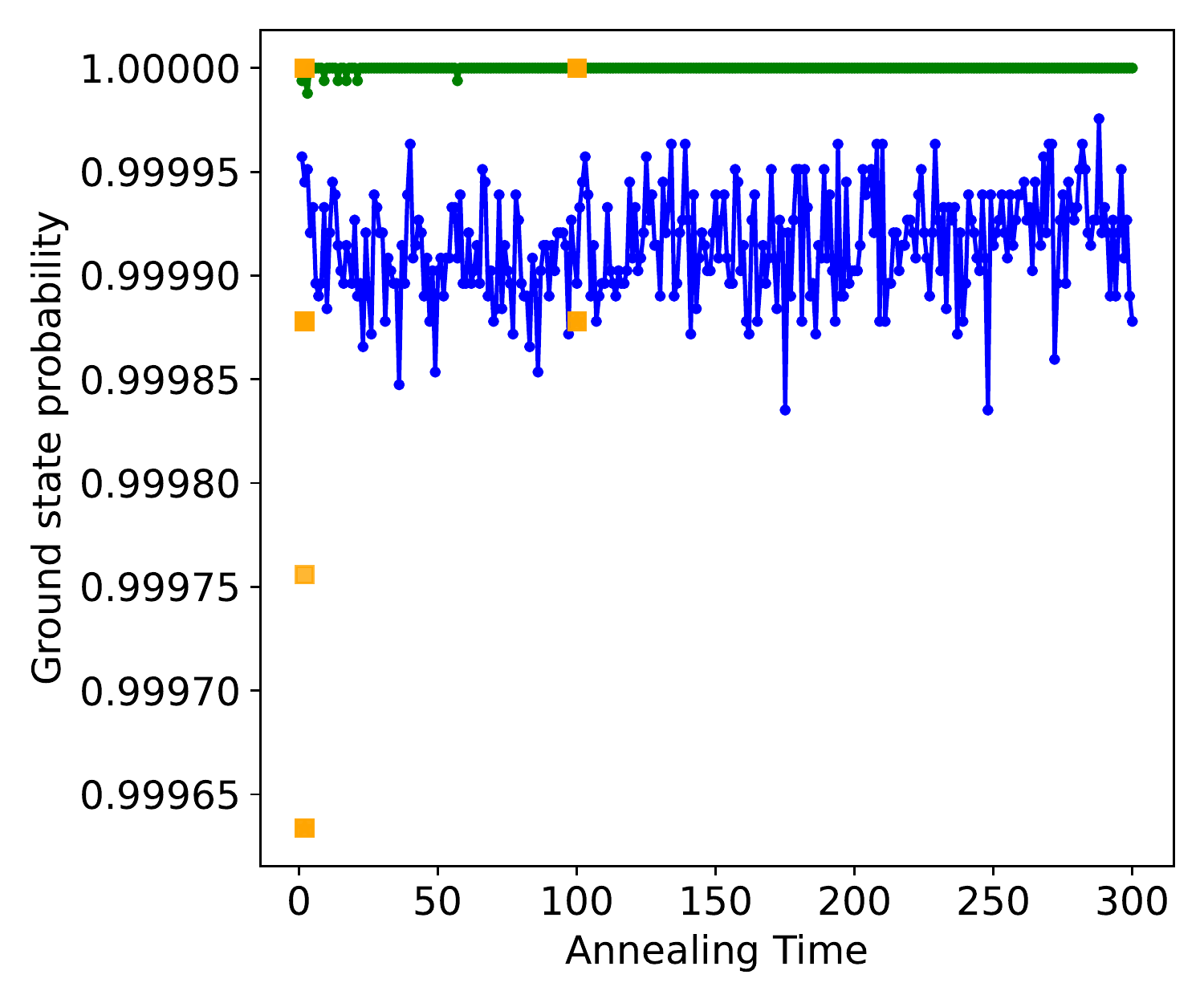}%
    \caption{Annealing-time dependent results on three D-Wave devices for Problems (a), (b), (c), (d), (e) (from left to right)
        for \textbf{(top)} log Fairness and \textbf{(bottom)} Ground state probability.
    }
    \label{fig:results_dwave}
\end{figure*}

Figure \ref{fig:results_GSP_vs_NSTR} directly shows a scatter plot of our two main metrics GSP vs. Fairness for all 12 NISQ backends using a logarithmic scale. There are several ways in which the D-Wave experiments were performed (see Section \ref{sec:methods_DWave}), resulting in several different results being plotted for this particular comparison. Note that for the gate model devices, we used 1 round QAOA. This means that we can not achieve the GSP on the gate model circuits (see \ref{tab:models}) that we can on the quantum annealers. Thus, in Figure \ref{fig:results_GSP_vs_NSTR} the GSP for all D-Wave backends is larger than the gate model devices. 

Ignoring individual outlier points, Figure \ref{fig:results_GSP_vs_NSTR} lets us visualize the Pareto front in this bi-criteria optimization, where a point lies on the Pareto front if it does not get outperformed in both metrics by other points. D-Wave excels at GSP, leaving gate-based models far behind in most cases. Recall that we could increase the theoretical GSPs for gate-based models by increasing the number of QAOA rounds, but that would inevitably lead to larger circuits with larger error and thus result in lower experimental GSP for gate models. In Problem (a), note the strong performance of IBMs QV16 Melbourne system; IonQ performs very well with several points on the Pareto front. In Problem (b), IonQ is the clear winner with IBMs Manhattan and Sydney system coming in second best in GSP and Fairness respectively. Problem (c) has a large GSP difference between D-Wave and all others and D-Wave also performs well in Fairness; among gate models, IonQ reaches high in Fairness. IBM~Q systems dominate in Problem (d) among gate models, interestingly the two QV32 backends Casablanca and Toronto outperform their QV128 sibling Montreal again. Problem (e) sees the gate models come close to D-Wave's GSP values, in particular for IonQ, which in turn gets outperformed in Fairness by most IBM~Q systems. 
Earlier work \cite{golden2021qaoa} has found that high error rates turn NISQ devices into random number generators, which actually makes them fair. The relatively good performance of IBM~Q Melbourne in Problem (e) and Rigetti in Problem (b) with respect to fairness are likely an example of this effect. 

In summary, Problems (a)-(e) show a remarkable diversity with no clear winner platform across all problems and a few surprises as described above. Rigetti does fall behind in most examples, except in GSP for Problem (c) and Fairness for Problem (b).

\subsection{Compilation options for IBM~Q backends}
\label{sec:results_ibm_compilation}

The particular QISKIT transpilation method used in order to compile each of the five circuits onto the IBM~Q backends greatly affects the results. In this Section, we present an analysis of how different transpilation options change the \textit{aggregate error}. 

Figure \ref{fig:results_IBM_transpilation}  (bottom row) shows the \textit{aggregate error} across all IBM~Q backends for each of the five test problems. In particular, we plot the \textit{aggregate error} for the three transpiler options outlined in Section \ref{sec:methods_IBMQ}. Additionally, we perform a grid search across the transpiler options \textit{layout-method}, \textit{routing-method}, \textit{translation-method}, and \textit{optimization-level}, and we plot the minimum and maximum aggregate error found in the grid search. 

This figure shows that there is no clear winning transpiler parameter combination across all backends and circuits, however the resulting aggregate varies by as much as 60 \%. In practice the figures suggest that taking the better performing of the noise adaptive compilation and giving initial layout compilation should result in near minimum aggregate error in most cases.

Figure \ref{fig:results_IBM_transpilation} (top row) plots \textit{aggregate error} as a function of IBM~Q backend for the 4 different possible QISKIT transpiler optimization levels. Besides the optimization flag, the \textit{layout-method} option was set to \textit{noise adaptive}. The figure shows that optimization level 3 performs best with the exception of the Sydney backend. Generally, increasing optimization levels decrease aggregate error as expected, but optimization level 1 is quite the exception to this rule.

\subsection{D-Wave}
\label{sec:results_dwave}

We investigate how our metrics change as a function of \textit{annealing time} for three D-Wave quantum annealing backends. For the LANL 2000Q backend, we plot the mean metric out of the 20 device calls (each call with 8192 shots) as a function of \textit{annealing time}. For the other two backends (Advantage-System1.1 and 2000Q-6), we plot all 20 points for the two annealing times they were run on (2 microseconds and 100 microseconds). 

In Figure \ref{fig:results_dwave} (top row), we observe a high variance in the fairness metric across all problems. Problems (a), (b) consistently show that the LANL 2000Q balanced embedding has the highest fairness over annealing time. Problems (c), (d), (e) on the other hand show that the LANL 2000Q minorminer embedding has the highest fairness over annealing time. 

In Figure \ref{fig:results_dwave} (bottom row), we observe different \textit{annealing time} dependent trends across each of the problems, largely from the LANL 2000Q balanced embedding data (the LANL 2000Q minorminer embedding results are relatively constant over annealing time). For Problems (a), (c), and (d), we observe a sharp increase in GSP in the first few microseconds of annealing time. Problems (a) and (d) then decrease in GSP after that initial increase. Problem (c) continues to increase GSP up until it plateaus at roughly 100 microseconds. Problem (b) shows no initial increase in GSP, instead it starts out at a high GSP and decreases over all annealing times. Problem (e) shows effectively no difference in GSP over time. It is hard to tell whether these D-Wave results are in agreement with the theoretical predictions of \cite{K_nz_2019}, which call for initial fairness with low annealing times and then unfairness but higher GSP as we increase annealing times, because all modern D-Wave backends have a minimum annealing time of 1 microsecond.

\section{Conclusion}

In this article, we compare IBM~Q's, Rigetti's, IonQ's and D-Wave NISQ devices with respect to their performance in finding ground states and fairness across a set of five paradigmatic problems. While no platform turned out to be dominant, we noticed performance differences in both metrics. It is encouraging to see that leading NISQ vendors are relatively close in performance and we hope that further technological advances will increase the fairness of sampling as well as bring advances to various compilation techniques. 

As for open problems, we plan to compare D-Wave performance against QAOA-based circuits that use the transverse field mixer (as opposed to the Grover mixer), we would like to relate our D-Wave results more cleanly to theoretical predictions. We also plan to continue our studies for novel NISQ backends.

\bibliographystyle{plainurl}
\bibliography{references}

\begin{thebibliography}{10}

\bibitem{akshay2020}
V~Akshay, H~Philathong, Mauro~ES Morales, and Jacob~D Biamonte.
\newblock Reachability deficits in quantum approximate optimization.
\newblock {\em Physical Review Letters}, 124(9):090504, 2020.
\newblock \href {http://arxiv.org/abs/1906.11259} {\path{arXiv:1906.11259}},
  \href {https://doi.org/10.1103/PhysRevLett.124.090504}
  {\path{doi:10.1103/PhysRevLett.124.090504}}.

\bibitem{azinovic2017assessment}
Marlon Azinovi{\'c}, Daniel Herr, Bettina Heim, Ethan Brown, and Matthias
  Troyer.
\newblock Assessment of quantum annealing for the construction of
  satisfiability filters.
\newblock {\em SciPost Physics}, 2(2):013, 2017.
\newblock \href {http://arxiv.org/abs/1607.03329} {\path{arXiv:1607.03329}},
  \href {https://doi.org/10.21468/SciPostPhys.2.2.013}
  {\path{doi:10.21468/SciPostPhys.2.2.013}}.

\bibitem{baertschi2020grover}
Andreas Bärtschi and Stephan Eidenbenz.
\newblock {Grover Mixers for QAOA: Shifting Complexity from Mixer Design to
  State Preparation}.
\newblock In {\em IEEE International Conference on Quantum Computing \&
  Engineering QCE'20}, pages 72--82, 2020.
\newblock \href {http://arxiv.org/abs/2006.00354} {\path{arXiv:2006.00354}},
  \href {https://doi.org/10.1109/QCE49297.2020.00020}
  {\path{doi:10.1109/QCE49297.2020.00020}}.

\bibitem{MM-paper}
Jun Cai, William~G. Macready, and Aidan Roy.
\newblock A practical heuristic for finding graph minors.
\newblock {\em arXiv e-prints}, 2014.
\newblock \href {http://arxiv.org/abs/1406.2741} {\path{arXiv:1406.2741}}.

\bibitem{MM-software}
{D-Wave~Systems}.
\newblock minorminer.
\newblock \url{https://github.com/dwavesystems/minorminer}, 2017.
\newblock A heuristic tool for minor embedding.

\bibitem{minorminer}
{D-Wave Systems}.
\newblock {\em
  \href{https://docs.ocean.dwavesys.com/projects/minorminer/en/latest/}{D-Wave
  Ocearn: minorminer}}, 2021.

\bibitem{golden2021qaoa}
John Golden, Andreas B{\"a}rtschi, Daniel O'Malley, and Stephan Eidenbenz.
\newblock {QAOA-based Fair Sampling on NISQ Devices}.
\newblock {\em arXiv e-prints}, 2021.
\newblock \href {http://arxiv.org/abs/2101.03258} {\path{arXiv:2101.03258}}.

\bibitem{grover2005fixed}
Lov~K. Grover.
\newblock {Fixed-Point Quantum Search}.
\newblock {\em Physical Review Letters}, 95(15):150501, 2005.
\newblock \href {https://doi.org/10.1103/PhysRevLett.95.150501}
  {\path{doi:10.1103/PhysRevLett.95.150501}}.

\bibitem{hadfield_qaoa}
Stuart Hadfield, Zhihui Wang, Bryan O’Gorman, Eleanor~G Rieffel, Davide
  Venturelli, and Rupak Biswas.
\newblock From the quantum approximate optimization algorithm to a quantum
  alternating operator ansatz.
\newblock {\em Algorithms}, 12(2):34, 2019.
\newblock \href {http://arxiv.org/abs/1709.03489} {\path{arXiv:1709.03489}},
  \href {https://doi.org/10.3390/a12020034} {\path{doi:10.3390/a12020034}}.

\bibitem{harp2008aquifer}
Dylan~R Harp, Zhenxue Dai, Andrew~V Wolfsberg, Jasper~A Vrugt, Bruce~A
  Robinson, and Velimir~V Vesselinov.
\newblock Aquifer structure identification using stochastic inversion.
\newblock {\em Geophysical Research Letters}, 35(8):L08404, 2008.
\newblock \href {https://doi.org/10.1029/2008GL033585}
  {\path{doi:10.1029/2008GL033585}}.

\bibitem{kadowaki1998quantum}
Tadashi Kadowaki and Hidetoshi Nishimori.
\newblock Quantum annealing in the transverse ising model.
\newblock {\em Physical Review E}, 58(5):5355, 1998.
\newblock \href {http://arxiv.org/abs/cond-mat/9804280}
  {\path{arXiv:cond-mat/9804280}}, \href
  {https://doi.org/10.1103/PhysRevE.58.5355}
  {\path{doi:10.1103/PhysRevE.58.5355}}.

\bibitem{kumar2020achieving}
Vaibhaw Kumar, Casey Tomlin, Curt Nehrkorn, Daniel O'Malley, and Joseph
  Dulny~III.
\newblock Achieving fair sampling in quantum annealing.
\newblock {\em arXiv e-prints}, 2020.
\newblock \href {http://arxiv.org/abs/2007.08487} {\path{arXiv:2007.08487}}.

\bibitem{K_nz_2019}
Mario~S. Könz, Guglielmo Mazzola, Andrew~J. Ochoa, Helmut~G. Katzgraber, and
  Matthias Troyer.
\newblock Uncertain fate of fair sampling in quantum annealing.
\newblock {\em Physical Review A}, 100(3):030303, 2019.
\newblock \href {http://arxiv.org/abs/1806.06081} {\path{arXiv:1806.06081}},
  \href {https://doi.org/10.1103/PhysRevA.100.030303}
  {\path{doi:10.1103/PhysRevA.100.030303}}.

\bibitem{mandra2017exponentially}
Salvatore Mandra, Zheng Zhu, and Helmut~G Katzgraber.
\newblock {Exponentially Biased Ground-State Sampling of Quantum Annealing
  Machines with Transverse-Field Driving Hamiltonians}.
\newblock {\em Physical Review Letters}, 118(7):070502, 2017.
\newblock \href {http://arxiv.org/abs/1606.07146} {\path{arXiv:1606.07146}},
  \href {https://doi.org/10.1103/PhysRevLett.118.070502}
  {\path{doi:10.1103/PhysRevLett.118.070502}}.

\bibitem{matsuda2009ground}
Yoshiki Matsuda, Hidetoshi Nishimori, and Helmut~G Katzgraber.
\newblock Ground-state statistics from annealing algorithms: quantum versus
  classical approaches.
\newblock {\em New Journal of Physics}, 11(7):073021, 2009.
\newblock \href {http://arxiv.org/abs/0808.0365} {\path{arXiv:0808.0365}},
  \href {https://doi.org/10.1088/1367-2630/11/7/073021}
  {\path{doi:10.1088/1367-2630/11/7/073021}}.

\bibitem{o2018approach}
Daniel O'Malley.
\newblock An approach to quantum-computational hydrologic inverse analysis.
\newblock {\em Scientific reports}, 8(1):1--9, 2018.
\newblock \href {https://doi.org/10.1038/s41598-018-25206-0}
  {\path{doi:10.1038/s41598-018-25206-0}}.

\bibitem{sieberer2018programmable}
Lukas~M Sieberer and Wolfgang Lechner.
\newblock {Programmable superpositions of Ising configurations}.
\newblock {\em Physical Review A}, 97(5):052329, 2018.
\newblock \href {http://arxiv.org/abs/1708.02533} {\path{arXiv:1708.02533}},
  \href {https://doi.org/10.1103/PhysRevA.97.052329}
  {\path{doi:10.1103/PhysRevA.97.052329}}.

\bibitem{yamamoto2020fair}
Masayuki Yamamoto, Masayuki Ohzeki, and Kazuyuki Tanaka.
\newblock {Fair Sampling by Simulated Annealing on Quantum Annealer}.
\newblock {\em Journal of the Physical Society of Japan}, 89(2):025002, 2020.
\newblock \href {http://arxiv.org/abs/1912.10701} {\path{arXiv:1912.10701}},
  \href {https://doi.org/10.7566/JPSJ.89.025002}
  {\path{doi:10.7566/JPSJ.89.025002}}.

\bibitem{yoder2014fixed}
Theodore~J Yoder, Guang~Hao Low, and Isaac~L Chuang.
\newblock {Fixed-Point Quantum Search with an Optimal Number of Queries}.
\newblock {\em Physical Review Letters}, 113(21):210501, 2014.
\newblock \href {http://arxiv.org/abs/1409.3305} {\path{arXiv:1409.3305}},
  \href {https://doi.org/10.1103/PhysRevLett.113.210501}
  {\path{doi:10.1103/PhysRevLett.113.210501}}.

\end{thebibliography}

\end{document}